\documentclass[twocolumn,superscriptaddress,nofootinbib,prb]{revtex4-1}

\usepackage{amsmath,graphicx,todonotes,multirow,array,subcaption}

\newcommand{\boldr}{{\bf r}}
\newcommand{\boldx}{{\bf x}}
\newcommand{\boldn}{{\bf n}}

\newcommand{\dr}{{\textrm d}{\bf r}}
\newcommand{\dx}{{\textrm d}{\bf x}}
\newcommand{\calf}{{\cal F}}

\usepackage[normalem]{ulem}

\begin{document}

\title{Influence of the fluid structure on the binding potential: comparing liquid drop profiles from density functional theory with results from mesoscopic theory}
\author{Adam P. Hughes}
\affiliation{Department of Mathematical Sciences, Loughborough University, Loughborough, LE11 3TU, UK}
\author{Uwe Thiele}
\affiliation{Institut f\"ur Theoretische Physik, Westf\"alische Wilhelms-Universit\"at M\"unster, Wilhelm Klemm Str.\ 9, 48149 M\"unster, Germany}
\affiliation{Center of Nonlinear Science (CeNoS), Westf{\"a}lische Wilhelms-Universit\"at M\"unster, Corrensstr.\ 2, 48149 M\"unster, Germany}
\affiliation{Center for Multiscale Theory and Computation (CMTC), Westf{\"a}lische Wilhelms-Universit\"at, Corrensstr.\ 40, 48149 M\"unster, Germany}
\author{Andrew J. Archer}
\affiliation{Department of Mathematical Sciences, Loughborough University, Loughborough, LE11 3TU, UK}
\date{\today}

\begin{abstract}
For a film of liquid on a solid surface, the binding potential $g(h)$ gives the free energy as a function of the film thickness $h$ and also the closely related structural disjoining pressure $\Pi = - \partial g / \partial h$. The wetting behaviour of the liquid is encoded in the binding potential and the equilibrium film thickness corresponds to the value at the minimum of $g(h)$. Here, the method we developed in [J.\ Chem.\ Phys.\ {\bf 142}, 074702 (2015)], and applied with a simple discrete lattice-gas model, is used with continuum density functional theory (DFT) to calculate the binding potential for a Lennard-Jones fluid and other simple liquids. The DFT used is based on fundamental measure theory and so incorporates the influence of the layered packing of molecules at the surface and the corresponding oscillatory density profile. The binding potential is frequently input in mesoscale models from which liquid drop shapes and even dynamics can be calculated. Here we show that the equilibrium droplet profiles calculated using the mesoscale theory are in good agreement with the profiles calculated directly from the microscopic DFT. For liquids composed of particles where the range of the attraction is much less than the diameter of the particles, we find that at low temperatures $g(h)$ decays in an oscillatory fashion with increasing $h$, leading to highly structured terraced liquid droplets.
\end{abstract}

\maketitle

\section{Introduction}
The wetting behaviour of liquids\cite{hansen13, schick90, degennes85, de2013capillarity} on a substrate is important in a great range of industrial and biological fields. The lubricating properties and the time it takes the liquid to evaporate to leave a dry surface are just two examples of properties that depend on the wetting behaviour of the liquid. A key quantity in the study of wetting is the binding, or interface, potential $g(h)$. It gives the contribution to the excess free energy of the system arising from having a film of liquid of thickness $h$ on a surface. The total grand potential for a system with volume $V$, containing a single planar solid surface with area $A$ covered by a liquid film of thickness $h$ is
\begin{equation}\label{eq:bp_def}
\Omega=-pV+A[\gamma_{wl}+\gamma_{lg} + g(h)+\Gamma\delta\mu],
\end{equation}
where $p$ is the pressure, $\gamma_{wl}$ is the wall-liquid interfacial tension, $\gamma_{lg}$ is the liquid-gas interfacial tension, $\Gamma\sim h$ is the adsorption on the surface and $\delta\mu=(\mu_{coex}-\mu)$. The function $g(h)${, which may be considered to be defined by Eq.\ \eqref{eq:bp_def} [cf.\ Eq.\ \eqref{eq:g_calc}],} has the property that $g(h)\to0$ as $h\to\infty$ when the chemical potential in the system $\mu$ is equal to the value at bulk gas-liquid phase coexistence $\mu_{coex}$. For finite film thickness $h$, $g(h)$ describes the contribution of the interaction between the two interfaces to the free energy and is related to the disjoining pressure $\Pi = - \partial g / \partial h$. 

For a given liquid in contact with a specific substrate, the binding potential $g(h)$ encodes important information about the wetting of the liquid on the substrate. The location of the global minimum of the binding potential indicates if the liquid wets the substrate or not. When the system is at gas-liquid phase coexistence, where $\delta\mu=0$, a global minimum at $h\to\infty$ indicates that the liquid does wet the surface -- i.e.\ that a drop placed on the surface spreads and that it is energetically favourable for the vapour to condense onto the surface to thicken the wetting film. On the other hand, when the global minimum is at a finite value of the film height, $h_0$, this indicates the liquid is partially wetting -- i.e.\ it does not completely wet the surface. $h_0$ is the equilibrium film thickness on the dry surface. This is the so-called ``precursor film'' thickness, although given that often this film can be sub-monolayer in thickness, this name might be considered misleading. The value of the binding potential at $h_0$ determines the equilibrium contact angle that a liquid-gas interface makes with the substrate: \cite{churaev95a,rauscher08}
\begin{equation}\label{eq:cang}
\theta=\cos^{-1}\left(1+\frac{g(h_0)}{\gamma_{lg}}\right).
\end{equation}
When considering the free energy of a system with a non-uniform thickness film of liquid on a surface, with film thickness profile $h(\boldx)$, where $\boldx=(x,y)$ is the coordinate of a point on the surface, the binding potential is then included into the widely used interfacial Hamiltonial (IH) model,\cite{dietrich88, schick90, macdowell02, macdowell13b, macdowell14} where the free energy of the system is given by the functional
\begin{equation}\label{eq:ih}
F[h] = \int \left[g(h) + \gamma_{lg} \sqrt{1+(\nabla h)^2} \right] \dx,
\end{equation}
{which is a generalisation of Eq.\ \eqref{eq:bp_def} to the inhomogeneous case and neglecting all terms that are constant or linear in $h$.} The second term in the integral is simply the surface area of the film multiplied by the gas-liquid interfacial tension. Minimising this functional gives the equilibrium shape of a liquid droplet. This free energy is often further approximated by assuming that the gradients in the film height profile are small, i.e.\ that $\sqrt{1+(\nabla h)^2}\approx1+\frac{1}{2}(\nabla h)^2$, which gives:
\begin{equation}
F[h] = \int \left[g(h) + \frac{\gamma_{lg}}{2} (\nabla h)^2 \right] \dx,
\label{eq:small_grad_F}
\end{equation}
where we have neglected an irrelevant constant term. Interestingly, this is the same energy functional that is minimised by the dynamical equations that one obtains from fluid mechanics. Starting from the Navier-Stokes equation to describe the flow of the film of liquid on the surface, together with kinematic and dynamic boundary conditions at the free surface and making also the small gradient (or long-wave) approximation, one obtains the thin film evolution {equation\cite{mitlin93, oron97, thiele07, thiele10}} which describes the dynamics of a thin liquid film:
\begin{equation}
\frac{\partial h}{\partial t} = \nabla \cdot \left[ Q(h) \nabla \frac{\delta F[h]}{\delta h} \right],
\label{eq:tfe}
\end{equation}
where the free energy functional $F$ is that in Eq.\ \eqref{eq:small_grad_F}. The most common approximation used for the mobility coefficient is $Q(h)=h^3/3\eta$, where $\eta$ is the viscosity of the liquid. {This is what is obtained in the long-wave approximation with a no-slip boundary condition at the surface. However, other approximations are also used, particularly if the effects of slip are important,\cite{oron97,MWW2005jem} or if transport by diffusion is incorporated.\cite{yin2016films,HLHT2015l}}

What the above discussion should make clear is that the binding potential $g(h)$ is a key fundamental quantity for describing the behaviour of liquids at surfaces. Expressions used for this quantity are often obtained via an asymptotic expansion,\cite{dietrich88, schick90, dietrich91} such as: $g(h)= a h^{-2} + b h^{-3} + \cdots $, valid when there are long-range (London dispersion) interactions between the fluid particles, or $g(h)=c \exp(-h/\xi) + d \exp(-2h/\xi) + \cdots$, when only short-range particle interactions are assumed. In both expressions, $a$, $b$, $c$, $d$ and $\xi$ are all parameters that depend on the state point of the system, i.e.\ on the value of the chemical potential $\mu$ and the temperature $T$. Approximations for $g(h)$ that are a combination of these two forms are also used.\cite{oron97, frastia12} These forms for $g(h)$ are asymptotic expressions, in principle only valid for large $h$. These should be expected to break down in the microscopic regime, as $h \to 0$.

In our previous paper, \cite{hughes15} we developed a density functional theory (DFT) based approach for calculating the binding potential, that is valid for all values of $h$. DFT is a microscopic statistical mechanical theory, which calculates the particle density distribution $\rho(\boldr)$ at all points $\boldr=(x,y,z)$ on and above the surface, and so this approach incorporates the influence of the inter-particle interactions down to the smallest relevant length scales. {Ref.\ \onlinecite{hughes15} also presents} results from applying the method to a simple discrete DFT for a lattice-gas (LG) model fluid. In order to validate the coarse-graining procedure for going from the microscopic density distribution level description, to the coarse grained film height profile $h(\boldx)$ level description, we calculated the density profile $\rho(\boldr)$ for liquid droplets on a surface and then compared the film height profiles $h(\boldx)$ obtained from Eq.\ \eqref{eq:ih}, together with the expression for $g(h)$ that we obtained from our DFT based method. {Overall the method works well, showing very good agreement between the two approaches. However, there are some aspects that are unsatisfactory and these relate to the fact that the LG is used to model the fluid. In particular, due to the discrete nature of the LG model, oscillations that do not decay in amplitude are sometimes present in the $h\to\infty$ tails of the binding potentials. Thus, while the LG model is adequate for the purpose of describing many aspects of fluids at interfaces and for demonstrating the validity of our method for calculating $g(h)$, the LG fluid does exhibit some non-physical discretisation artefacts.} These inaccuracies motivate the research described here, which utilises a continuum DFT model. Here, we apply the approach of Ref.\ \onlinecite{hughes15} to calculate the binding potential for a model Lennard-Jones (LJ) fluid at various different model substrates using a state-of-the-art continuum DFT. We also calculate the density profile for droplets on surfaces and compare the film height profiles obtained from the density profiles with the film height profiles obtained from minimising Eq.\ \eqref{eq:ih} together with the binding potentials obtained from our DFT based method. We find good agreement, further validating our coarse-graining method. Note that a similar `parameter-passing' is done in Ref.~\onlinecite{tretyakov13} where binding potentials are obtained via microscopic molecular-dynamics computer simulations. Refs.\ \onlinecite{md2006jcp, MacDowell2011, MacDowell2014adcolintsci, md2015pre} also present binding potentials obtained via computer simulations.

We also study the effect on the binding potential and drop profiles of having a short-ranged attraction between the liquid particles. In particular, we consider the Asakura-Oosawa model\cite{asakura54} for colloid-polymer mixtures. For this system we find oscillatory binding potentials that lead to striking terraced drop height profiles.

This paper is laid out as follows: In Sec.~\ref{sec:model} the used DFT model is introduced and in Sec.~\ref{sec:constraint} the method for calculating density profiles with a specified value of the adsorption is described. From the resulting profiles, the binding potential is calculated and results from this are displayed in Sec.~\ref{sec:binding_pot}. This section also contains a comparison of droplet height profiles calculated directly using DFT with the height profiles from the IH [Eq.~\eqref{eq:ih}], together with the binding potentials calculated using DFT. In Sec.~\ref{cont:sec:fw}, results for the colloid-polymer mixture which has oscillatory binding potentials and forms terraced droplet profiles are presented. We finish with a brief summary and conclusions in Sec.~\ref{sec:conc}.

\section{DFT for the fluid}\label{sec:model}

We consider a fluid of particles interacting via the pair potential $u(r)$, which is strongly repulsive for small distances $r$ between the particles, and then at larger $r$ is attractive. This potential may be split into a repulsive part $u_\textrm{r}(r)$ and an attractive part $v(r)$, which sum together to give the full potential $u(r)=u_\textrm{r}(r)+v(r)$. Thus, in the usual manner for simple liquids,\cite{hansen13} we may use a perturbative approximation for the excess Helmholtz free energy
\begin{equation}\label{cont:vdw}
F_\textrm{ex}[\rho(\boldr)] = F_\textrm{r}[\rho(\boldr)] + \frac{1}{2} \iint \rho(\boldr) \rho(\boldr') v(| \boldr - \boldr'| ) \dr \dr',
\end{equation}
which is a functional of the one-body density profile $\rho(\boldr)$ and where $F_\textrm{r}[\rho]$ is the excess Helmholtz free energy for the reference fluid of particles interacting via the potential $u_\textrm{r}(r)$. As long as a judicious splitting of the potential is made,\cite{hansen13,barker76} then $F_\textrm{r}[\rho]\approx F_\textrm{hs}[\rho]$, where $F_\textrm{hs}[\rho]$ is the excess Helmholtz free energy for a hard-sphere fluid; here we use the highly accurate White Bear version of fundamental measure theory (FMT).\cite{hansen13,roth02} In FMT the excess free energy is expressed as
\begin{equation}
\calf_\textrm{hs} = \int \Phi(\{n_\alpha\}) \dr,
\end{equation}
where 
\begin{align}
\Phi = & -n_0 \ln(1-n_3) + \frac{n_1n_2- \boldn_1 \cdot \boldn_2}{1-n_3} \nonumber \\
& + (n_2^3-3n_2 \boldn_2 \cdot \boldn_2) \frac{n_3+(1-n_3)^2\ln(1-n_3)}{36\pi n_3^2(1-n_3)^2},
\end{align}
is a function of a set of weighted densities $\{n_\alpha\}$ which are convolutions of the fluid density profile and a set of weight functions based on the geometrical measures of a sphere:
\begin{equation}
n_\alpha(\boldr) = \int \rho(\boldr') w_\alpha(|\boldr-\boldr'|) \dr'.
\end{equation}
The set of four scalar and two vector weight functions, $\{w_\alpha\}$, is defined as
\begin{align}
w_3(\boldr) & = \theta(R-r), \nonumber \\
w_2(\boldr) & = \delta(R-r), \nonumber \\
w_1(\boldr) & = \frac{w_2(\boldr)}{4 \pi R}, \nonumber \\
w_0(\boldr) & = \frac{w_2(\boldr)}{4 \pi R^2}, \nonumber \\
{\bf w}_2(\boldr) & = \frac{\boldr}{r} \delta(R-r), \nonumber \\
{\bf w}_1(\boldr) & = \frac{{\bf w}_2(\boldr)}{4 \pi R}, \label{eq:weights}
\end{align}
where $r=|\boldr|$, $R$ is the hard-sphere radius, $\theta(r)$ is the Heaviside step function and $\delta(r)$ is the Dirac delta function. For more details, a good review of FMT can be found in Ref.\,\onlinecite{roth10}.

The total grand free energy of the system is
\begin{eqnarray}\label{eq:gfe}
\Omega[\rho(\boldr)]  = &k_BT \int \rho(\boldr) (\ln [\Lambda^3 \rho(\boldr)] - 1) \dr+ F_\textrm{ex}[\rho(\boldr)]\notag \\
	&  + \int \rho(\boldr) \left( V(\boldr) - \mu \right)\dr,
\end{eqnarray}
where the first term is the ideal-gas contribution to the free energy, with $k_B$ being Boltzmann's constant, $T$ the temperature and $\Lambda$ the thermal de Broglie wavelength. In the last term, $V(\boldr)$ is the external potential and $\mu$ is the chemical potential. The equilibrium fluid density profile is that which minimises Eq.\,\eqref{eq:gfe} and so is the solution to the Euler-Lagrange equation
\begin{equation}\label{eq:el}
\frac{\delta \Omega}{\delta \rho} = k_BT \ln(\Lambda^3 \rho) +\frac{\delta F_\textrm{ex}}{\delta \rho} + V(\boldr) - \mu = 0,
\end{equation}
which can be solved iteratively via Picard iteration on the equation
\begin{equation}\label{eq:pic}
\rho(\boldr) = \rho_b \exp\left[c^{(1)}(\boldr) - \beta V(\boldr) - c^\infty \right],
\end{equation}
which is obtained from Eq.\ \eqref{eq:el} and where $\beta=(k_BT)^{-1}$ is the inverse temperature. The quantity $c^{(1)}(\boldr) = -\beta \delta F_\textrm{ex} / \delta \rho$ is the one-body direct correlation function and $c^\infty$ is defined $c^{(1)}(\boldr\to\infty) = c^\infty$, which assumes that the external field decays to zero in the bulk, where the fluid density is $\rho_b$. A detailed account of the use of Picard iteration can be found in Ref.~\onlinecite{roth10}. The key idea is that one starts from an initial guess for the density profile $\rho_{\rm old}$, which is substituted into the right hand side of Eq.\ \eqref{eq:pic}, to give the profile $\rho_{\rm rhs}$. A new guess for the density profile is obtained by mixing, $\rho_{\rm new}=m\rho_{\rm rhs}+(1-m)\rho_{\rm old}$, where the mixing parameter $m$ is typically in the range $0.005-0.1$. This is then repeated until convergence is achieved. 

We initially consider the case where the attractive pair interaction is that for a truncated and shifted Lennard-Jones (LJ) potential
\begin{equation}\label{cont:ljts}
v(r) = v_\textrm{LJ}(r) - v_\textrm{LJ}(r_c),
\end{equation}
where $r_c$ is a cut off range and
\begin{equation}\label{cont:3dpairpot}
v_\textrm{LJ}(r) = 
\begin{cases}
4 \epsilon \left( \left(\sigma/r\right)^{12} - \left(\sigma/r \right)^6 \right) & \mbox{ if } \sigma < r < r_c \\
0 & \mbox{ otherwise}.
\end{cases}
\end{equation}
$\sigma=2R$ is the diameter of the particles. The effect of varying the cut-off range $r_c$ is discussed below. The presence of attractive interactions means that gas-liquid phase coexistence can occur; indeed all of the results presented here are calculated at the point of liquid-gas coexistence.

\begin{figure}
\includegraphics[width=\columnwidth]{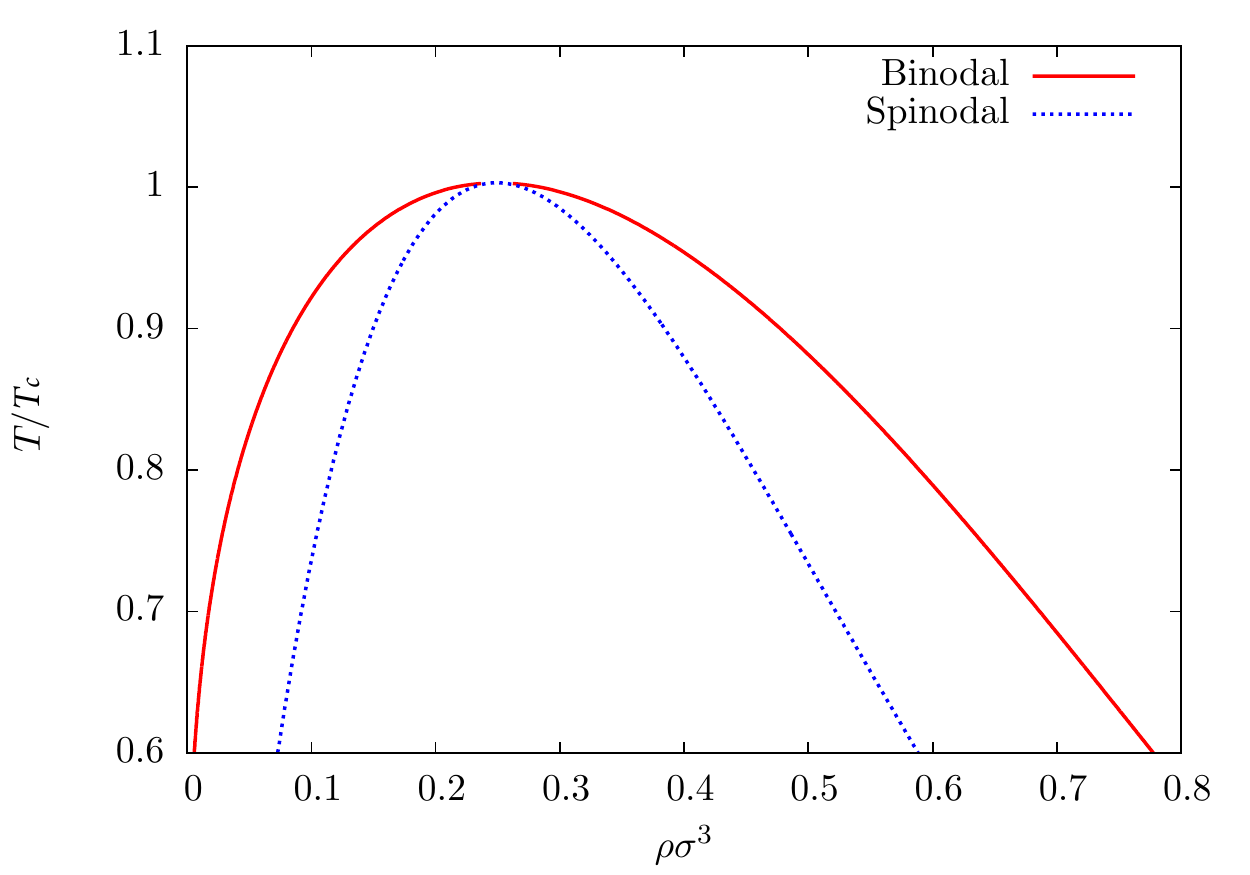}
\caption{Phase diagram of the fluid in the temperature $T$ versus density $\rho$ plane, where $T_c$ is the temperature at the critical point. The red solid curve shows the binodal, which are the coexisting liquid-gas densities. The blue dotted curve is the spinodal.}
\label{cont:fig:phase}
\end{figure}

In the uniform bulk fluid, the FMT weighted densities simplify; the vector weighted contributions vanish and the scalar weighted densities are simply the bulk density multiplied by the integrated weight function. The Helmholtz free energy per unit volume in the bulk fluid is
\begin{eqnarray}\label{cont:energypervol}
a(\rho) = k_BT \rho (\ln (\Lambda^3 \rho) - 1) + \frac{4\pi R^3 \rho^2}{1-\frac43 \pi R^3\rho}  \notag \\
+ \frac{\frac43 \pi R^3 \rho^2}{(1-\frac43 \pi R^3\rho)^2}
+ \frac{1}{2} \rho^2 \int v(\boldr) \dr.
\end{eqnarray}
From this we can take derivatives to obtain the pressure $p$ and the chemical potential $\mu$.\cite{hansen13} The bulk fluid phase diagram in Fig.\,\ref{cont:fig:phase} shows the coexisting liquid and gas densities, $\rho_l$ and $\rho_g$, which are found by solving the simultaneous equations
\begin{eqnarray}\label{cont:phase}
p(\rho_g) & = p(\rho_l), \notag \\
\mu(\rho_g) & = \mu(\rho_l),
\end{eqnarray}
at a fixed temperature $T$. This gives the points where the pressure, chemical potential and temperature are equal for each phase. These meet at the critical point at the temperature $T_c=1.21\epsilon/k_B$ and density $\rho_c\sigma^3=0.249$, for $r_c\to\infty$. The spinodal, which is the curve where $\partial^2 a/\partial\rho^2=0$, is also plotted in this phase diagram. 

Note that in Eq.\,\eqref{cont:energypervol} the form of $v(r)$ does not influence the value of $a(\rho)$, it is only the integrated value $\int v(\boldr) \dr\propto\epsilon$ which determines the free energy per unit volume. As such, the phase diagram of a fluid where the Helmholtz free energy is given by Eq.\,\eqref{cont:vdw} with any choice of $v(r)$ is just a rescaling of the curve in Fig.\,\ref{cont:fig:phase}. To ensure that varying the cut-off range $r_c$ does not alter the bulk fluid phase behaviour and phase diagram, the interaction strength $\epsilon$ is renormalised so that the integrated strength of the pair potential remains equal to the value when $r_c \to \infty$. In what follows, when an interaction strength $\epsilon$ is quoted, this is the equivalent value for $r_c \to \infty$ and the true interaction strength is
\begin{equation}\label{cont:renorm}
\epsilon_\textrm{true} = \epsilon / \left(1 + 2\left( \frac{\sigma}{r_c}\right)^9 - 3 \left(\frac{\sigma}{r_c}\right)^3 \right).
\end{equation}

We assume that the solid planar wall surface onto which the liquid is deposited is composed of particles interacting with the fluid particles via the pair potential Eq.\,\eqref{cont:ljts} and an additional hard-core repulsion term. Assuming too that the particles in the surface have a uniform density $\rho_w\sigma^3=f$ in the region $z<0$ and density equal to zero for $z\geq0$, then the net potential between the whole solid and a single fluid particle is
\begin{equation}\label{eq:extpot}
V(\boldr) = \frac{2\pi f \epsilon}{3} \left( \frac{2}{15} \left(\frac{\sigma}{z+\sigma/2}\right)^9 - \left(\frac{\sigma}{z+\sigma/2}\right)^3\right),
\end{equation}
for $z \geq \sigma/2$ and $V(\boldr)=V(x,y,z)=\infty$ for $z < \sigma/2$. Note that the product $f \epsilon$ determines the attractive strength of the wall and can be replaced by the single parameter $\epsilon_w = f \epsilon$. However, using the product notation $f \epsilon$ allows for easier comparison with the simulation data in Ref.~\onlinecite{ingebrigtsen07} and thus the parameter $f$ gives the relative attractive strength of the wall-fluid interactions compared to the fluid-fluid interactions. The absolute parameter $\epsilon_w$ allows an easier comparison of fluids defined by various choices of $\epsilon$ on the same substrate.

One motivation for considering the system with pair potential \eqref{cont:ljts} is that it is frequently used in simulations.\cite{tretyakov13,macdowell06,ingebrigtsen07} For the fluid at walls with three different values of $f$ we display in Table\,\ref{cont:tab:dft-sim} the values of the surface tensions and contact angles obtained using molecular dynamics (MD) simulations (from Ref.~\onlinecite{ingebrigtsen07}) and also from the present DFT. The agreement of the contact angles between the two approaches is surprisingly good. DFT is a mean field theory and so does not capture all of the interfacial fluctuations that {are present in the MD simulations} and so some discrepancy should be expected. The MD results are for the temperature $T=0.75T_c$, where $T_c$ is the critical temperature, and with $r_c=2.5\sigma$. The DFT results are equivalently for $\epsilon=(4/3)\epsilon_c$ where $\epsilon_c$ is the critical value of $\epsilon$ for a fixed $\beta$. This gives coexisting densities of $\rho_g \sigma^3=0.0127$ and $\rho_l \sigma^3 =0.7606$ in the  MD simulations\cite{ingebrigtsen07} and $\rho_g \sigma^3=0.0277$ and $\rho_l \sigma^3 = 0.6367$ from the present DFT. There are differences between the surface tensions but interestingly, the calculated contact angles are in very good agreement.

\begin{table}
\begin{tabular}{>{$}c<{$} c >{$}c<{$} >{$}c<{$} >{$}c<{$} >{$}c<{$}}
f & Model & \beta \gamma_{lg} & \beta \gamma_{wl} & \beta \gamma_{wg} & \theta \\
\hline
\multirow{2}{2em}{0.3} & MD & 0.489 & 0.375 & -0.014 & 137^\circ \\
	& DFT & 0.373 & 0.290 & -0.002 & 142^\circ \\
\multirow{2}{2em}{0.6} & MD & 0.489 & 0.028 & -0.014 & 99^\circ \\
	& DFT & 0.373 & 0.053 & -0.028 & 103^\circ \\
\multirow{2}{2em}{1.0} & MD & 0.489 & -0.548 & -0.062 & 39^\circ \\
	& DFT & 0.373 & -0.419 & -0.102 & 32^\circ
\end{tabular}
\caption{A comparison of contact angles and surface tensions obtained from molecular dynamics (MD) simulations {(from Ref.~\onlinecite{ingebrigtsen07}) with the present DFT model for the temperature $T=0.75T_c$ and $r_c=2.5\sigma$}.}
\label{cont:tab:dft-sim}
\end{table}

The interfacial tensions are excess free energies per unit area, so are straightforward to calculate using DFT. {From these, together with Young's equation
\begin{equation}
\gamma_{wg} = \gamma_{wl} + \gamma_{lg}\cos \theta,
\label{eq:Young}
\end{equation}
the contact angle $\theta$ can be obtained}. For example, to calculate the wall-liquid interfacial tension $\gamma_{wl}$, the equilibrium density profile of the fluid within the external field Eq.\,\eqref{eq:extpot} is found by solving Eq.~\eqref{eq:pic} for $\mu=\mu_{coex}$, the value of the chemical potential at bulk liquid-gas coexistence, with the density in bulk $\rho(z\to\infty)=\rho_l$. Substituting the resulting density profile into Eq.\ \eqref{eq:gfe} gives the grand potential of the system with the liquid at the wall, $\Omega_{wl}$. Then, the interfacial tension is obtained by subtracting the bulk contribution, $\gamma_{wl}=(\Omega_{wl}+pV)/A$, where $p$ is the bulk fluid pressure, $V$ is the volume of the system and $A$ is the area of the wall. Recall that for a bulk system the grand potential $\Omega=-pV$.

\begin{figure}
\includegraphics[width=\columnwidth]{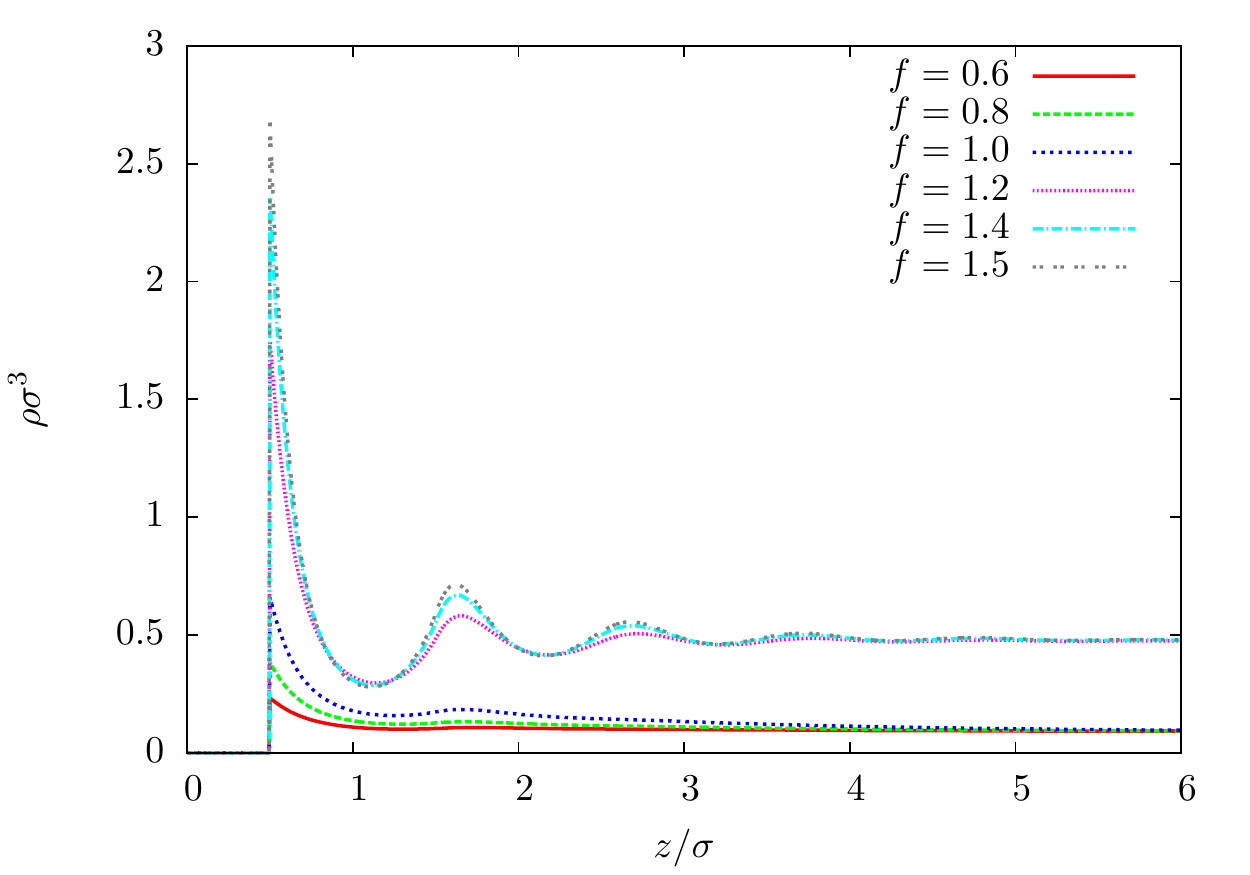}
\caption{Density profiles for the fluid with $\beta \epsilon = 1.1${, $r_c=5\sigma$} and $\mu=\mu_{coex}$ at a planar wall with varying attraction strength $f$. As $f$ is increased, the density at the wall increases and there is increased oscillatory structure in the profile, particularly for {$f > 1.06$}, when the liquid wets the wall and so there is a thick (macroscopic) film of the liquid adsorbed on the wall.}
\label{cont:fig:1dprof}
\end{figure}

The wall-gas interfacial tension $\gamma_{wg}$ is found in a similar manner, although the wall attraction strength parameter has to be sufficiently small that the liquid does not wet the wall. A selection of such density profiles are displayed in Fig.\,\ref{cont:fig:1dprof}. These are calculated for $\mu=\mu_{coex}$, with the density in bulk $\rho(z\to\infty)=\rho_g$. We see that for the smaller values of the wall attraction strength parameter {$f<1.06$}, there is a low gas-like density at the wall. However, a wetting film is observed for the curves corresponding to  the higher values of {$f>1.06$}. In these cases there is a thick film of the higher density liquid adsorbed on the wall. For {$f<1.06$} we obtain $\gamma_{wg}=(\Omega_{wg}+pV)/A$, where $\Omega_{wg}$ is the grand potential of the system with the gas in contact with the wall.

The liquid-gas surface tension $\gamma_{lg}$ is obtained by setting the external potential $V(\boldr)=0$ and then calculating the density profile for the free gas-liquid interface.\cite{evans79,evans92} Substituting the resulting profile back into Eq.~\eqref{eq:gfe} gives the total free energy of the system containing the interface and then the interfacial tension is given as $\gamma_{lg}=(\Omega_{lg}+pV)/A$, where $A$ is the area of the liquid-gas interface.

\section{Density profiles with adsorption constraint}\label{sec:constraint}

The procedure for calculating a binding potential using DFT is presented in Ref.\,\onlinecite{hughes15}, but is briefly summarised here. The binding potential $g(h)$ is calculated as a constrained free energy -- see Eq.~\eqref{eq:bp_def}. The constraint is that there is a specified thickness, $h$, of liquid on the wall. At the microscopic scale, particularly when the system is described by DFT, it is more natural to describe the liquid on the surface in terms of the adsorption $\Gamma$, rather than in terms of a film height $h$. The adsorption is defined as the excess fluid density per unit area
\begin{equation}
\Gamma = \frac{1}{A}\int (\rho(\boldr) - \rho_b) \dr.
\end{equation}
When the adsorbed film thickness is large enough then the adsorption is obviously related to the film height, since $\Gamma \approx h (\rho_l - \rho_g)$. In fact, one can define an effective film thickness as
\begin{equation}\label{eq:adsheight}
h\equiv\frac{\Gamma}{\rho_l-\rho_g},
\end{equation}
although it should be borne in mind that for small adsorption $\Gamma$, the effective film height $h$ can be much less than the diameter of the individual molecules. In the case of a depletion of 
the gas phase at a non-wetting wall, the adsorption and therefore effective film height may even become negative. Then evolution equations like (\ref{eq:tfe}) may only be used with an adequately adapted mobility $Q(h)$.

To calculate the constrained free energy $g(h)$, instead of specifying the value of $h$, we specify the value of the adsorption $\Gamma$; i.e.\ we calculate $g(\Gamma)$. Minimising the grand potential functional \eqref{eq:gfe}, together with Eq.\,\eqref{eq:extpot}, yields the equilibrium density profile with adsorption $\Gamma_0$, which when substituted back into  Eq.\ \eqref{eq:gfe} gives the minimum value of the binding potential $g(\Gamma_0)$ [c.f.~Eq.~\eqref{eq:cang}]. To find $g(\Gamma)$ for any other value of $\Gamma$, the system must be constrained. The full curve $g(\Gamma)$ is obtained by calculating for a series of values of $\Gamma$. At each value, the constrained fluid density profile is calculated and inserted into Eq.\ \eqref{eq:gfe} to obtain $\Omega(\Gamma)$. The binding potential is then
\begin{equation}\label{eq:g_calc}
g(\Gamma)=\frac{\Omega(\Gamma)+pV}{A}-\gamma_{lg}-\gamma_{wl}.
\end{equation}
The constrained density profile with specified $\Gamma$ is calculated using the normalisation method developed in Ref.\,\onlinecite{archer11}. In the Picard iteration of Eq.\,\eqref{eq:pic}, at each iteration the density profile is also normalised via
\begin{equation}\label{eq:ads}
\rho_\textrm{new} = (\rho_\textrm{old} - \rho_g) \frac{\Gamma}{\Gamma_\textrm{old}} + \rho_b,
\end{equation}
where $\Gamma_\textrm{old}$ is the adsorption of the current iterate density profile and $\Gamma$ is the target value for the adsorption. Note that normalising in this way on the {\em excess} density results in a profile that has the correct density value, $\rho_g$, away from the wall at large $z$. When the iteration converges, it yields a density profile with the desired value of $\Gamma$. This procedure in effect amounts to adjusting the Euler-Lagrange equation, Eq.\,\eqref{eq:el}, to 
\begin{equation}
\frac{\delta \Omega}{\delta \rho} = k_BT \ln(\Lambda^3 \rho) +\frac{\delta F_\textrm{ex}}{\delta \rho} + V(\boldr) + V_f(\boldr) - \mu = 0,
\end{equation}
where $V_f$ is an additional fictitious external potential, self-consistently calculated by the algorithm, that stabilises the density profile at the desired adsorption. It has the property $V_f(z\to\infty)=0$, so that the bulk gas density at $z\to\infty$ has the correct value, $\rho_g$. See Refs.\,\onlinecite{archer11, hughes15} for further discussion of this method for calculating the binding potential and the properties of $V_f$.

The values of the binding potential obtained over a range of $\Gamma$ can be fitted with a suitable choice of `fit function' to enable it to be used in the interfacial free energy \eqref{eq:ih}. For the LG model the fit function
\begin{equation}\label{eq:fit}
g(\Gamma) = \mathcal{A} \frac{\exp[- P(\Gamma)] -1}{\Gamma^2},
\end{equation}
with
\begin{equation}
P(\Gamma) = a_0 \Gamma^2 e^{-a_1 \Gamma} +\sum_{n=2}^{m} a_n \Gamma^n
\end{equation}
often gives an excellent fit over the whole range of values of $\Gamma$, as long as the wall-fluid dispersion interactions are not truncated.\cite{hughes15} $\mathcal{A}$ and $a_0, a_1, \dots$ are parameters to be fitted. The polynomial in $P(\Gamma)$ is normally truncated at $m=5$ or 6. If desired, via Eq.\,\eqref{eq:adsheight} this can also be expressed as a function of $h$ instead of $\Gamma$. For large $\Gamma$, Eq.\ \eqref{eq:fit} yields the correct $\sim\Gamma^{-2}$ decay, and for thin films with small $\Gamma$, Eq.\ \eqref{eq:fit} yields a polynomial plus an additional exponential from the first term in $P(\Gamma)$ that allows one to describe correctly the asymmetry of the minimum near $\Gamma\approx0$ that is present when the fluid does not wet the wall.\cite{hughes15} For many (but not all) of the results presented below, we find that Eq.\ \eqref{eq:fit} gives an excellent fit to the DFT data.

\begin{figure}

\includegraphics[width=0.95\columnwidth]{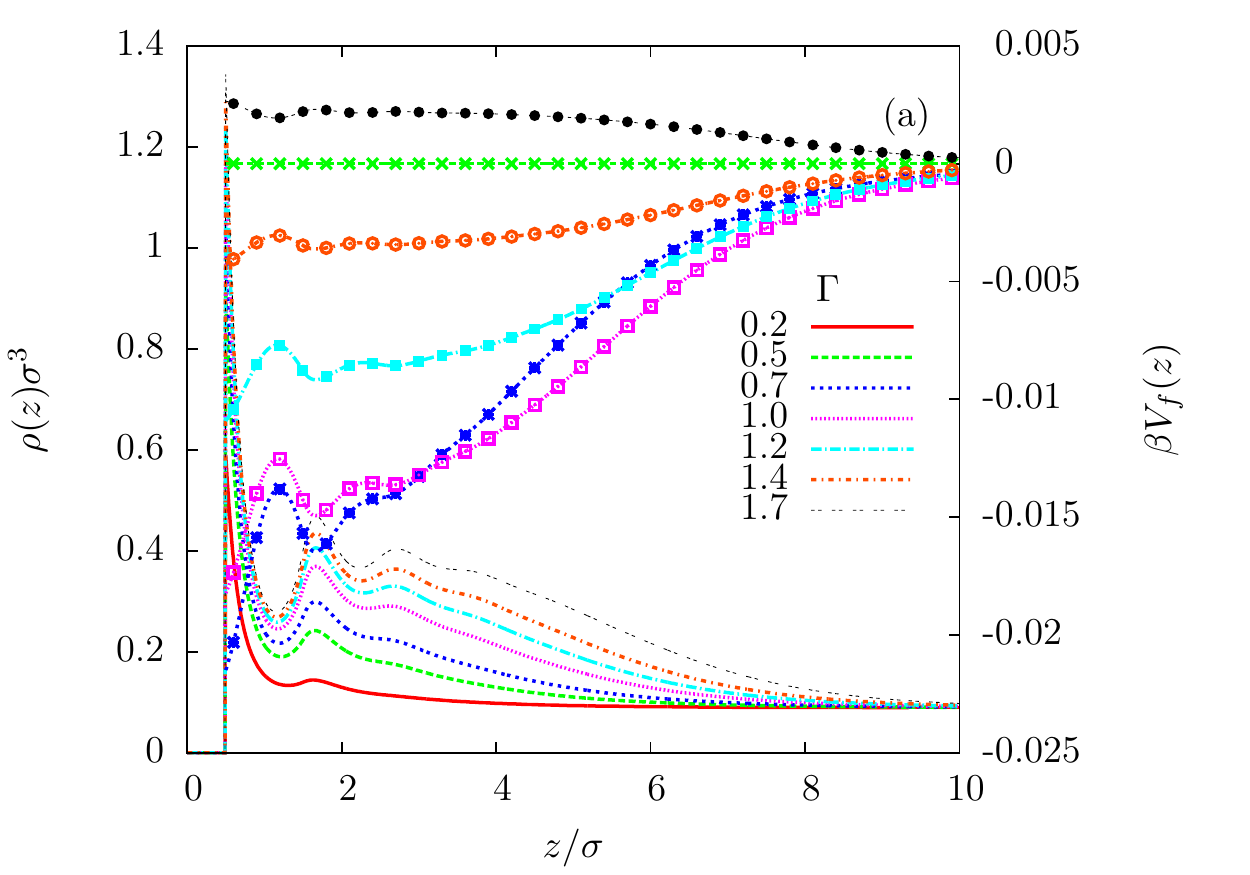}

\includegraphics[width=0.95\columnwidth]{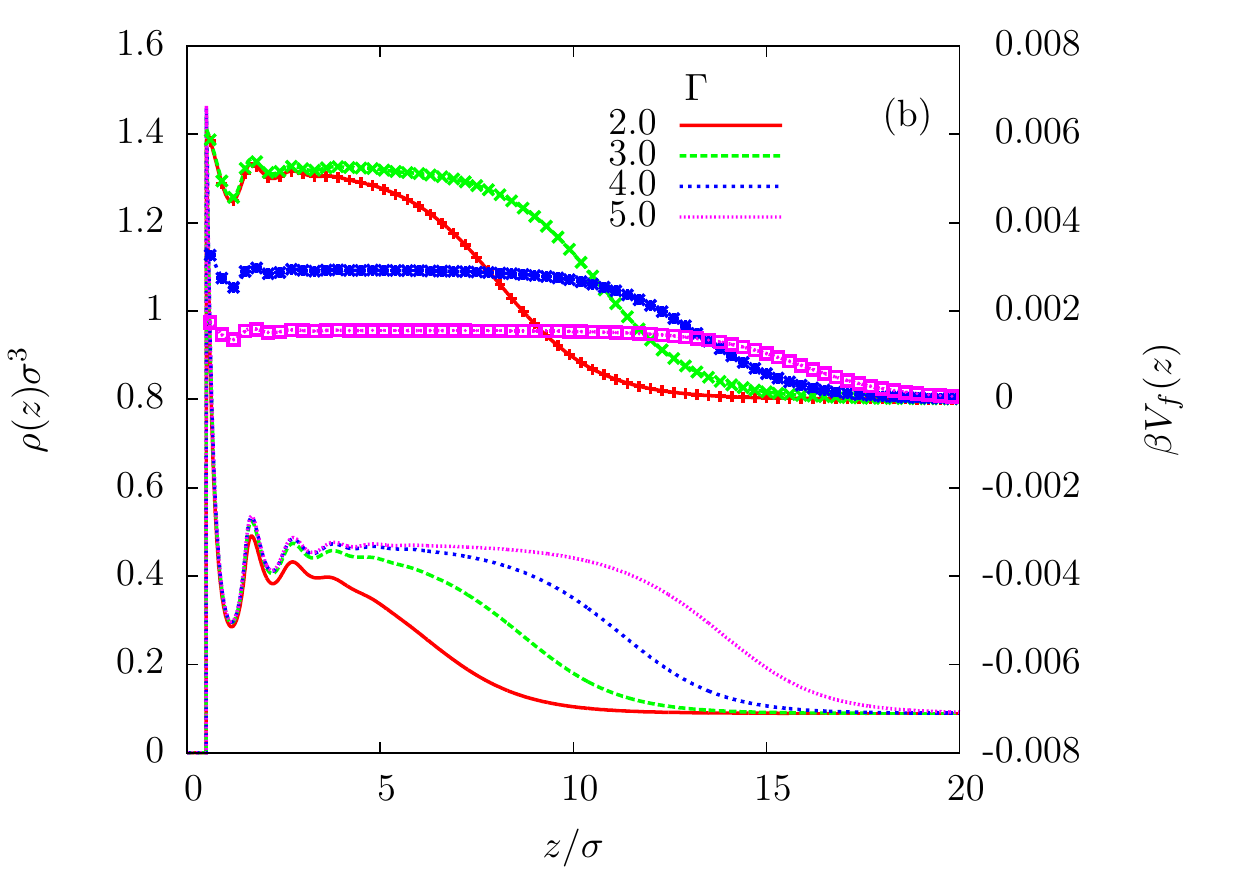}

\includegraphics[width=0.85\columnwidth]{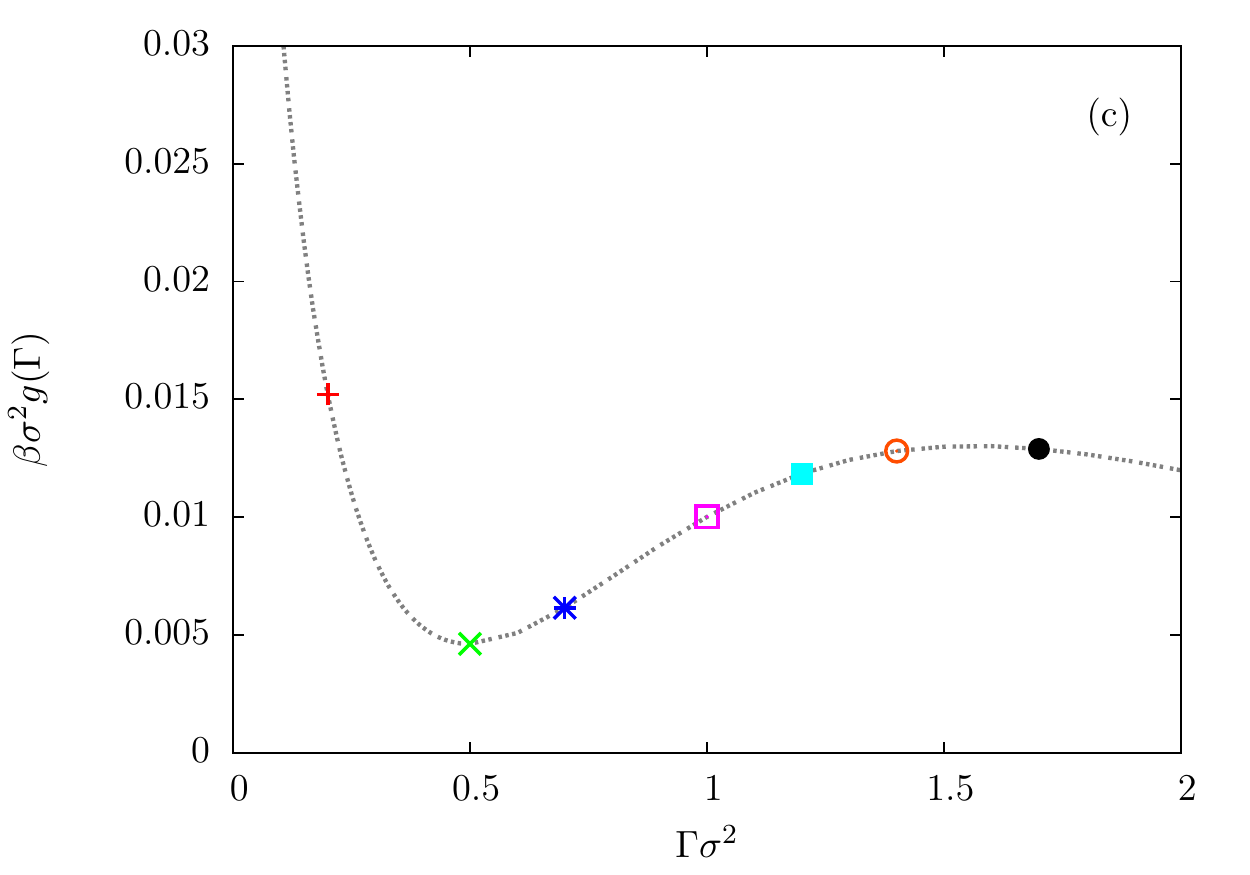}

\caption[Stabilising fictitious potentials]{Density profiles for a fluid with $\beta \epsilon=1.1$, {$r_c=5\sigma$} and $\mu=\mu_{coex}$ at the wall with attraction strength {$f=1.1$} with specified values of the adsorption displayed together with the fictitious external potentials that are needed to stabilise them. In (a) and (b) the plain lines are the density profiles (left axis) and the lines with symbols are the fictitious potentials (right axis). The fictitious potential for $\Gamma \sigma^2 =0.2$ is not displayed as its magnitude is greater than the scale displayed, it is approximately 100 times larger than the other potentials. Smaller adsorption values are shown in (a) which also correspond to the coloured symbols of the binding potential shown in (c). Note that in all the plots above only part of the system is displayed; the full system is of length $82\sigma$.}
\label{cont:fig:constrained1d} 
\end{figure}

For a series of specified values of $\Gamma$, in Fig.\,\ref{cont:fig:constrained1d} we display the corresponding density profiles $\rho(z)$ together with the fictitious potentials $V_f(z)$ required to stabilise them against the wall with attraction strength $f=1.0$. The lines without symbols are the density profiles (left axis scale) while the corresponding lines with symbols are the fictitious potentials (right axis scale). Density profiles with smaller values of $\Gamma$ are displayed in Fig.~\ref{cont:fig:constrained1d}(a), while those for larger $\Gamma$ are in Fig.~\ref{cont:fig:constrained1d}(b). In the profiles for larger $\Gamma$ we see near the wall a region where the density is almost uniform with a value equal to that of the coexisting bulk liquid. At larger $z$ there is an interface, beyond which the profiles decay to the bulk gas density value.

In Fig.~\ref{cont:fig:constrained1d}(c) we display the binding potential calculated from this series of density profiles via Eq.~\eqref{eq:g_calc}. The symbols on $g(\Gamma)$ refer to the results in Fig.~\ref{cont:fig:constrained1d}(a) with the corresponding point style. From the local gradient of $g(\Gamma)$ we can see why the fictitious potentials in (a) are either attractive or repulsive, or in the case of the profile with $\Gamma\sigma^2=0.5$, which is at the minimum of $g(\Gamma)$, we find $V_f(z)=0$. Consider, for example, the density profile with $\Gamma \sigma^2=1$. In this case $V_f$ is attractive and the gradient of $g(\Gamma)$ at this point is negative. If one were to remove the constraint and set $V_f=0$, then the system would relax down the gradient in $g(\Gamma)$ to the equilibrium density profile at $\Gamma \sigma^2=0.5$, decreasing the adsorption. Conversely, to stabilise a density profile with an adsorption value that lies outside of the potential well in $g(\Gamma)$, an additional repulsive external field is required. The unconstrained behaviour in this case would increase the adsorbed film thickness to reach the minimum at $\Gamma\to\infty$ (the global minimum in this case). The repulsive $V_f$ forces a lower adsorption. For the larger adsorption cases shown in Fig.~\ref{cont:fig:constrained1d}(b), the potential $V_f$ extends only as far as the liquid layer and the magnitude of $V_f$ decreases for larger adsorptions as the energetic minimum in $g(\Gamma)$ at $\Gamma\to\infty$ is approached. 

The density profile for $\Gamma\sigma^2=0.2$ (red solid line) is plotted in Fig.~\ref{cont:fig:constrained1d}(a) but the corresponding $V_f$ is not displayed because it is very large, approximately two orders of magnitude greater than the fictitious potentials for the other adsorption values displayed in Fig.\,\ref{cont:fig:constrained1d}. That the potential $V_f$ is large when the adsorption is very small is not surprising since to enforce $\Gamma \to 0$, one has to remove all of the liquid from the surface thereby cancelling the effect of the true external field that is attracting the fluid to the wall. Therefore, one must expect $V_f$ to be of the same magnitude as the true potential $V$ as $\Gamma \to 0$.

\begin{figure}
\centering

\includegraphics[width=0.44\columnwidth]{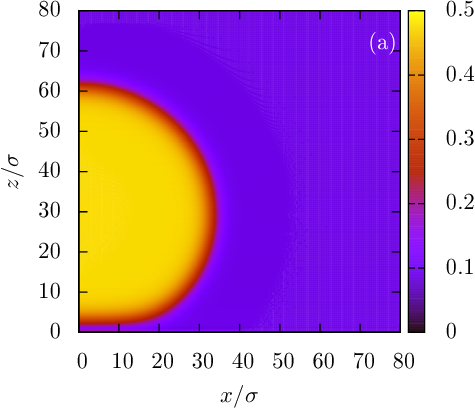}
\includegraphics[width=0.54\columnwidth]{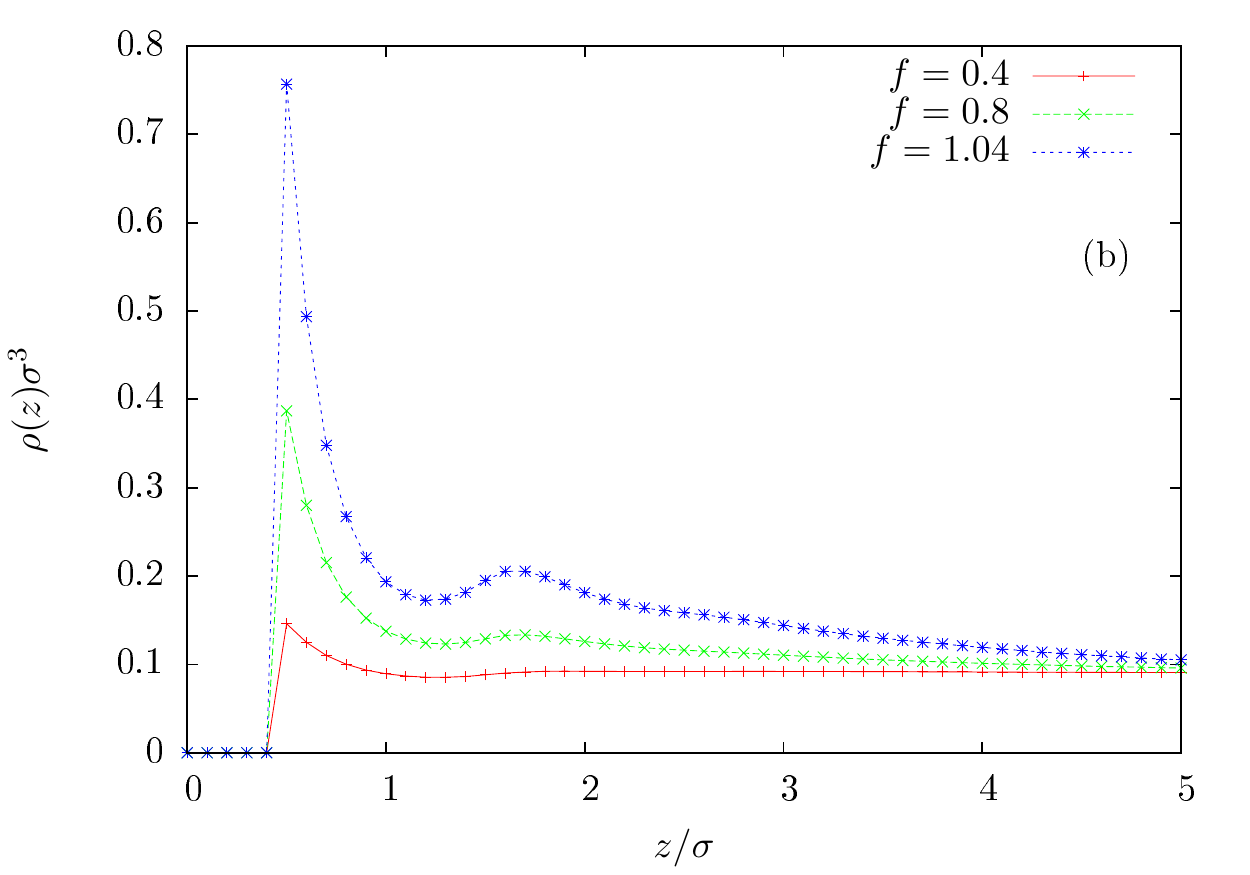}

\includegraphics[width=0.96\columnwidth]{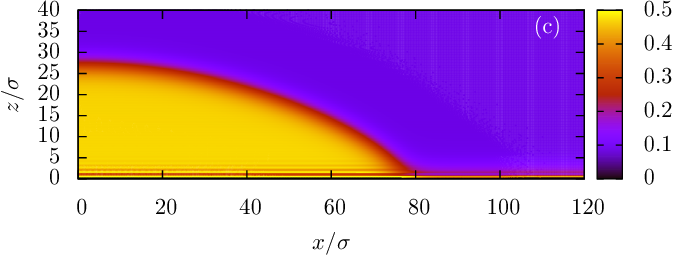}

\includegraphics[width=0.44\columnwidth]{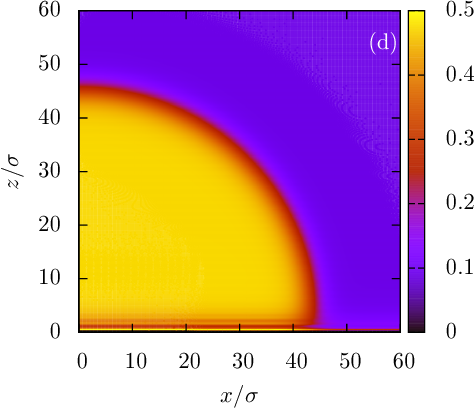}
\includegraphics[width=0.54\columnwidth]{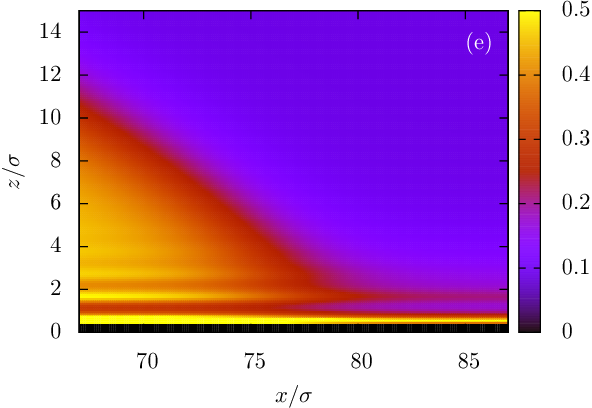}

\caption{Droplet profiles for a fluid of fixed volume with $\beta \epsilon = 1.1$ and $r_c=5${$\sigma$} against substrates of varying attraction strength $f$. The values of $f$ and the contact angles for the droplets in (a), (c) and (d) are $(f,\theta)= (0.4,151^\circ)$, $(1.04,20^\circ)$ and $(0.8,87^\circ)$, respectively. A plot of the density profiles at the wall in regions away from the droplets is shown in (b) and a zoom of the contact region of the droplet in (c) is shown in (e).}
\label{cont:fig:2dprof}
\end{figure}

In Fig.\,\ref{cont:fig:2dprof} we display density profiles calculated using the DFT assuming that the density profile can vary in the $x$ and $z$ directions, but is invariant in the $y$-direction, along the surface. With the same adsorption constraint applied, two-dimensional (2D) droplet profiles (i.e.\ three-dimensional ridges) can be found, as constrained equilibrium solutions. The selection of droplet profiles shown in Fig.\,\ref{cont:fig:2dprof} are for $f=0.4$, 1.04 and 0.8, which (see Fig.\ \ref{cont:fig:contactAngles} below) correspond to contact angles of $\theta= 151^\circ$, $20^\circ$ and $87^\circ$. There is significant structure in the density profiles in the vicinity of the wall, especially in the cases with smaller contact angle. The contact region where the edge of the droplet meets the substrate is quite diffuse with substantial density oscillations, as seen in Fig.~\ref{cont:fig:2dprof}(e). The effect of packing of particles in the body of the droplet can also be seen from the density peaks close to the substrate. The {thin adsorption layer outside the drop} is also present which shows just one or two density peaks. A zoom of the density profile in this region is shown in Fig.\,\ref{cont:fig:2dprof}(b) for all three droplets. For previous examples of droplet density profiles calculated using DFT, see Refs.~\onlinecite{berim08, ruckenstein10, pereira11, nold11, nold14, Giacomello19012016}.

\section{Binding potentials}\label{sec:binding_pot}

\begin{figure}
\includegraphics[width=\columnwidth]{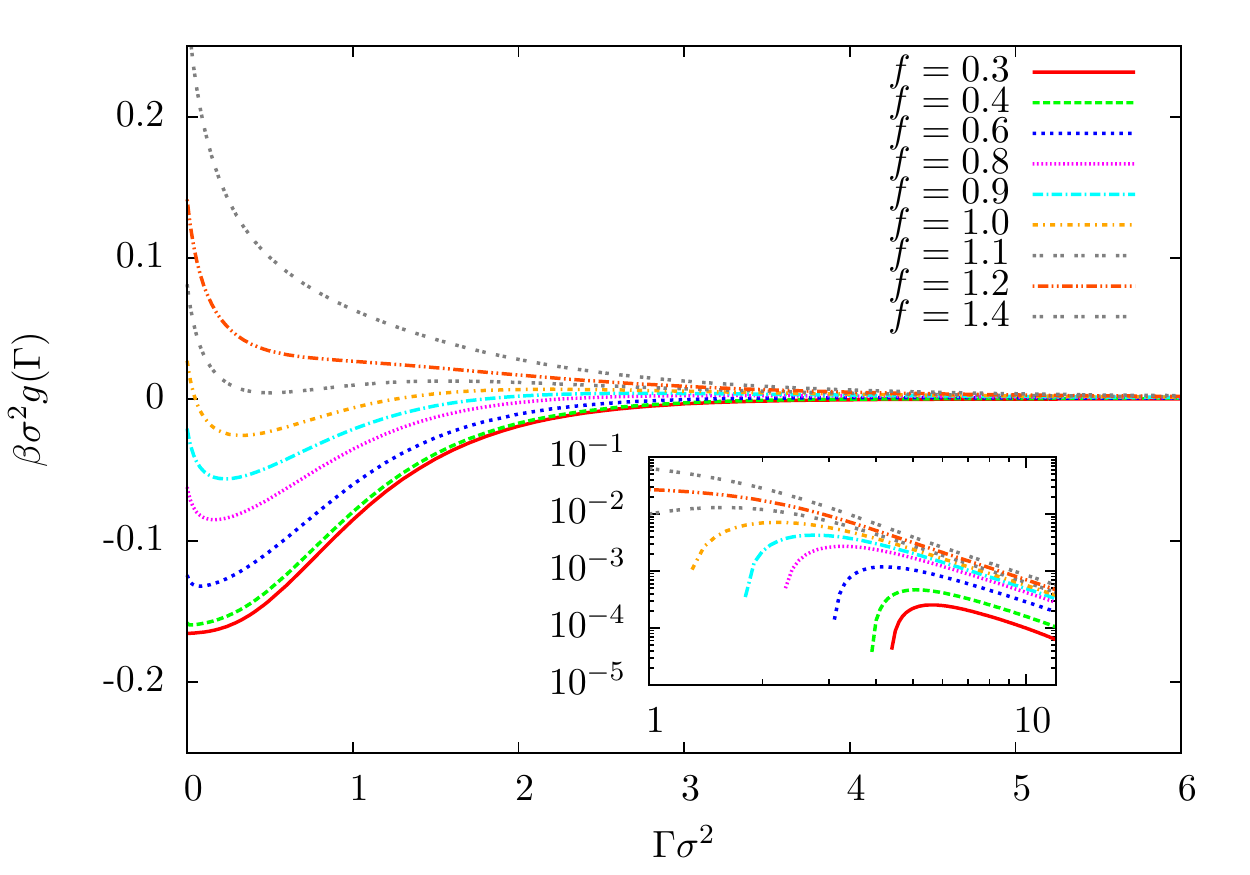}
\caption[Continuum binding potentials]{The binding potential $g(\Gamma)$ calculated using DFT for the fluid with $\beta \epsilon=1.1$ and pair-potential truncated beyond a range of $r_c=5\sigma$, for a series of different values of the wall attraction strength $f$, as indicated in the key. The inset shows the same data on a logarithmic scale. The binding potentials have the same $\sim\Gamma^{-2}$ decay, due to the external potential \eqref{eq:extpot} that is not truncated.}
\label{cont:fig:bind}
\end{figure}

In Fig.\,\ref{cont:fig:bind} we display the binding potentials calculated via Eq.~\eqref{eq:g_calc} for various values of the wall attraction strength parameter $f$, for a fluid with $\beta\epsilon=1.1$ and $r_c=5\sigma$. We see that for large values of $f$ the global minimum in $g(\Gamma)$ is at $\Gamma\to\infty$, corresponding to the case where the fluid wets the wall and the unconstrained equilibrium corresponds to a thick adsorbed film. As the wall attraction $f$ is decreased, a local minimum appears in $g(\Gamma)$ at a small finite value of $\Gamma$. This minimum then deepens, to become the global minimum at the wetting transition at {$f\approx1.06$}. For {$f<1.06$} the unconstrained equilibrium corresponds to a finite thickness film of liquid on the wall, with adsorption $\Gamma$ equal to the value at the minimum of $g(\Gamma)$.

The inset of Fig.\,\ref{cont:fig:bind} shows that the binding potentials exhibit the expected $\sim\Gamma^{-2}$ decay due to the power-law tail for $z\to\infty$ in the wall potential \eqref{eq:extpot}. The decay for $\Gamma\to\infty$ is monotonic, without oscillations; below in Sec.~\ref{cont:sec:fw} we present binding potentials that do have oscillatory decay. The binding potentials in Fig.\,\ref{cont:fig:bind} are qualitatively similar to those found in Ref.~\onlinecite{hughes15} for the LG model. One difference between the present continuum DFT results and the results for the LG is that the minimum in $g(\Gamma)$ that can be present for small values of $\Gamma$ occurs at lower values of the adsorption than in the LG model. In fact, when the wall is very weakly attractive (e.g.~for $f=0.3$) then the minimum in $g(\Gamma)$ occurs at a slightly negative value of $\Gamma$.

These binding potentials should  be related to the density profiles that are displayed in Fig.\,\ref{cont:fig:1dprof}. Note that for {$f<1.06$} the global minimum occurs at a low value of the adsorption which corresponds to a non-wetting fluid. This is reflected in the density profiles of Fig.\,\ref{cont:fig:1dprof} where the corresponding density profiles all have only a small amount of fluid adsorbed to the wall. The global energetic minimum occurs at infinite adsorption in Fig.\,\ref{cont:fig:bind} for the external potentials with $f=1.2$ and $f=1.4$ and then density profiles for these cases in Fig.\,\ref{cont:fig:1dprof} show a macroscopic liquid layer adsorbed to the wall.

\begin{figure}
\includegraphics[width=0.95\columnwidth]{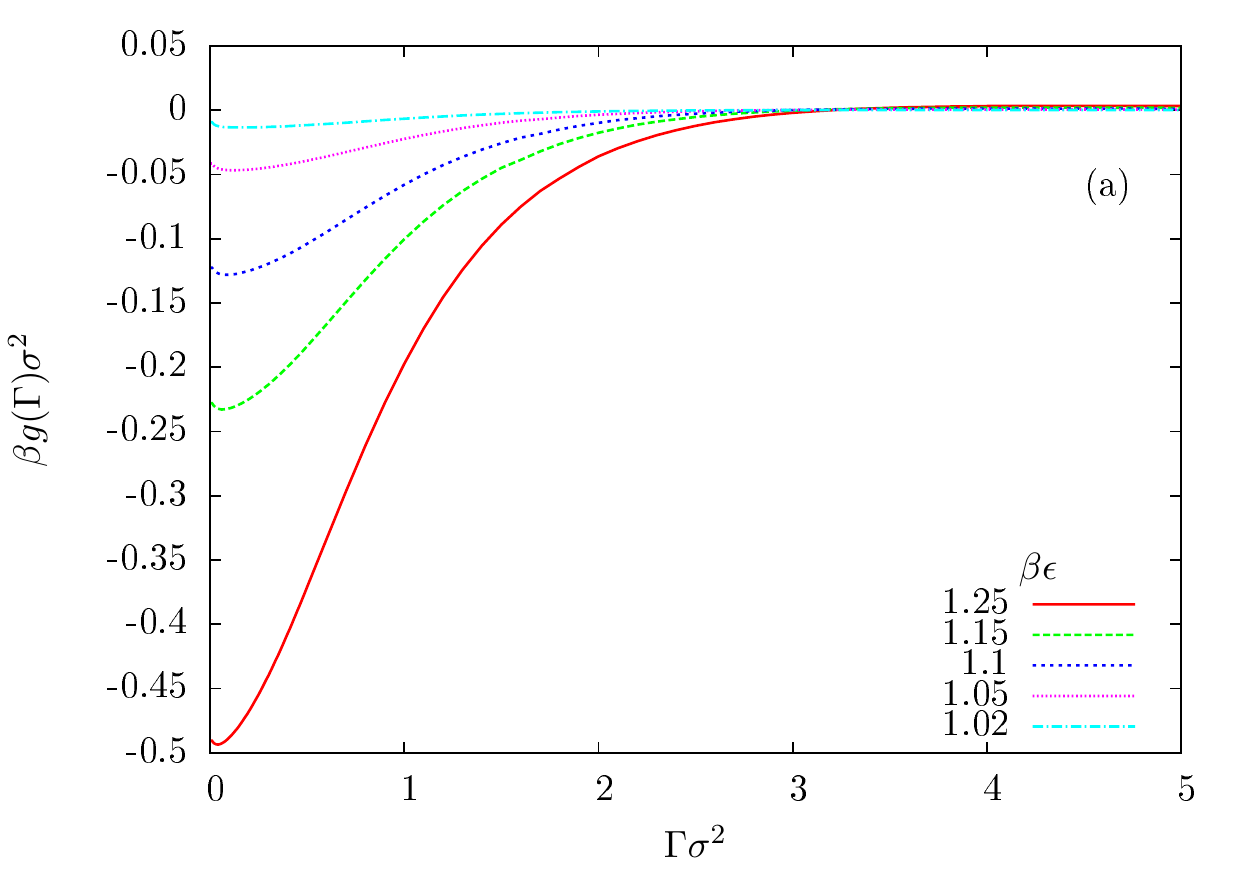}

\includegraphics[width=0.95\columnwidth]{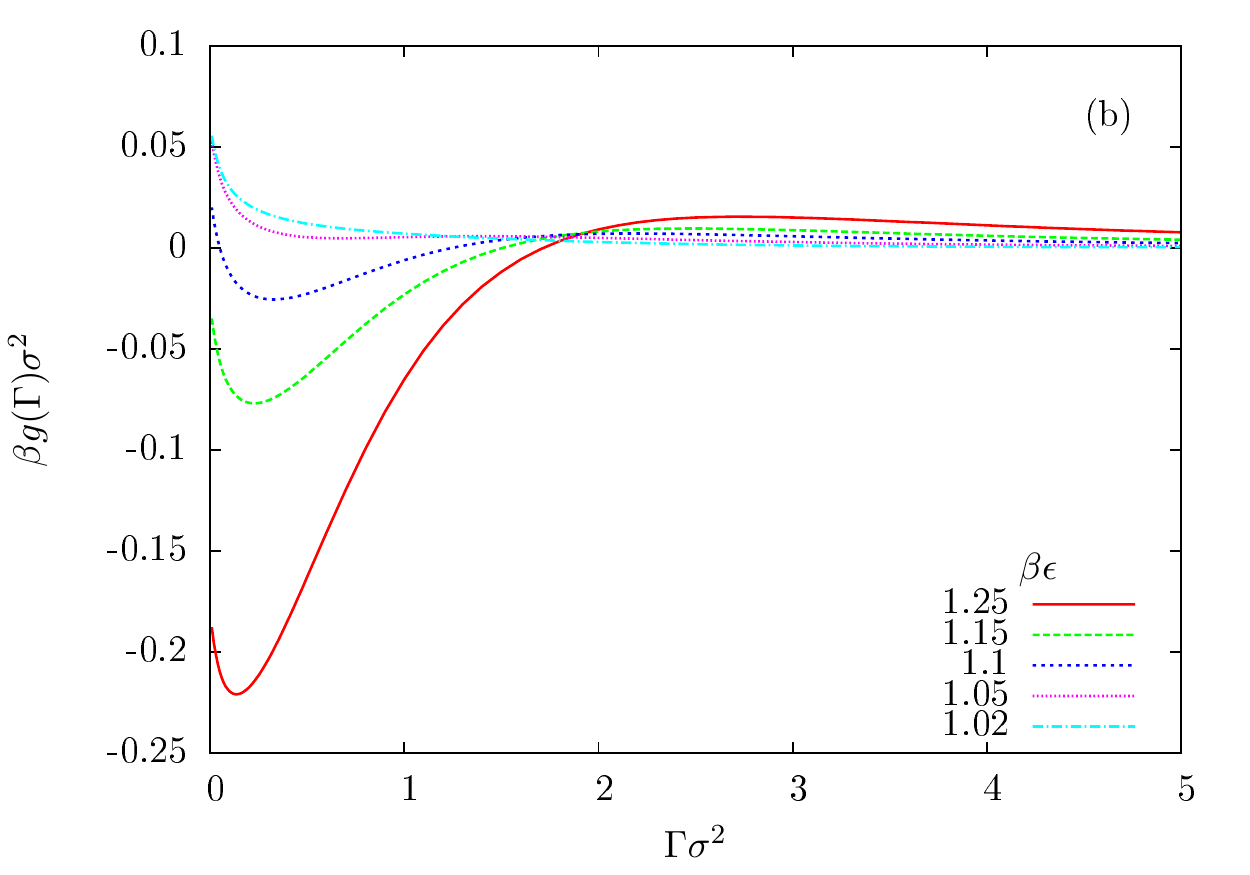}
\caption[Change in minimum value of the binding potential]{As the strength of fluid-fluid interactions is increased (or equivalently the temperature is decreased) the fluid becomes less wetting and the minimum in $g(\Gamma)$ becomes deeper. Results are shown for (a) a substrate of fixed attraction strength $\beta\epsilon_w=0.7$ and (b) a substrate with attraction strength equal to the fluid-fluid interaction strength, i.e.\ fixed $f=1$.}
\label{cont:fig:epsBind}
\end{figure}

The depth of the potential well in $g(\Gamma)$ is strongly dependent on the fluid-fluid interaction strength $\epsilon$, or equivalently, on the temperature. As the temperature is decreased (or $\epsilon$ is increased) below the bulk critical temperature, the densities of the coexisting liquid and gas phases move further apart and the potential well in the binding potential becomes deeper. This behaviour is illustrated in Fig.\,\ref{cont:fig:epsBind}. The case in (a) is for fixed $\beta \epsilon_w = 0.7$. The external potential does not change as the value of $\beta\epsilon$ is varied. For the case displayed in Fig.\,\ref{cont:fig:epsBind}(b), the strength of the wall attraction is defined relative to the fluid-fluid interactions, i.e.\ $\beta \epsilon_w = f \beta\epsilon$, so that the attractive strength of the substrate is increased with increasing $\beta\epsilon$. Nevertheless, increasing $\beta\epsilon$ still leads to the minimum in $g(\Gamma)$ becoming deeper. Independent of how the external potential is chosen, the depth of the potential well at small $\Gamma$ decreases as the bulk fluid critical temperature is approached.

\begin{figure}
\includegraphics[width=\columnwidth]{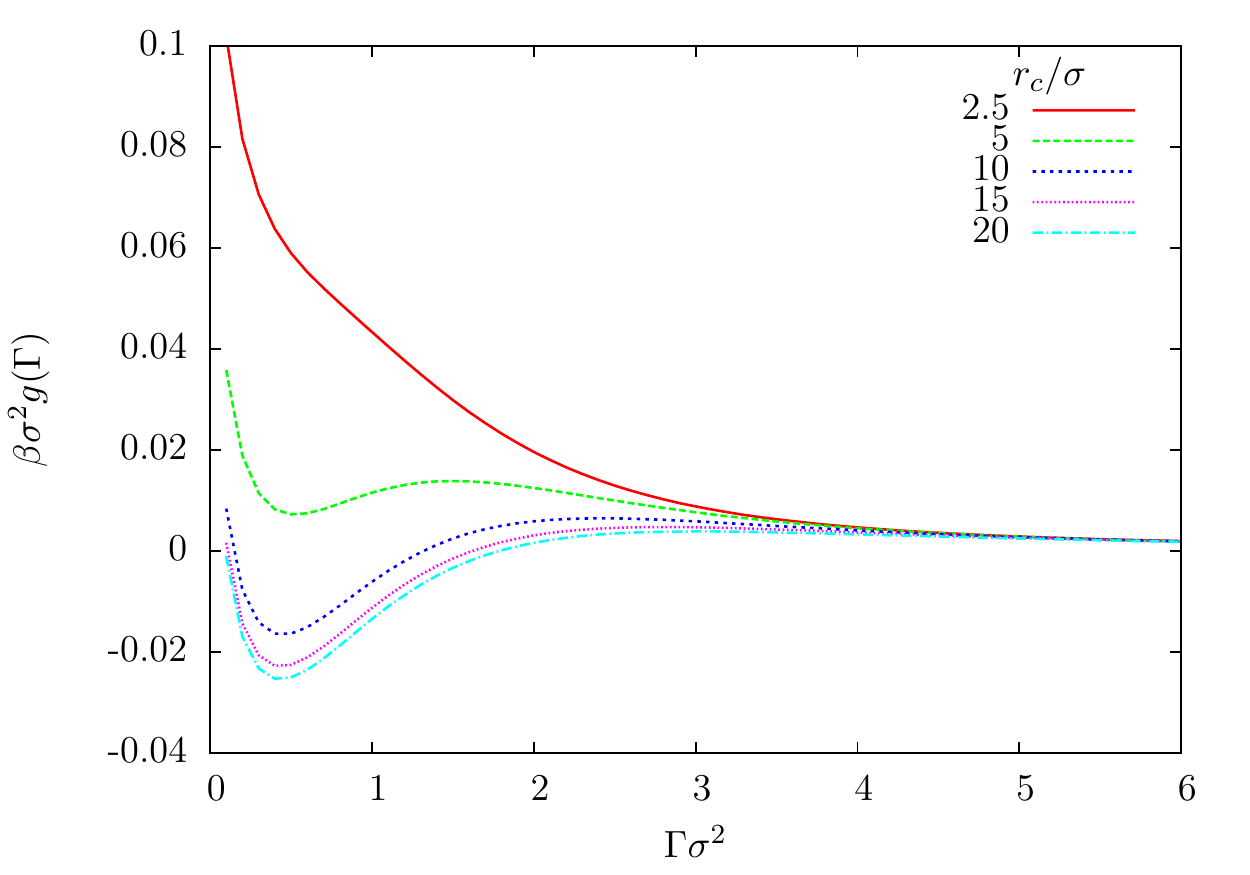}
\caption[Effect of truncating interaction ranges]{The binding potential for $\beta \epsilon=1.1$ and $\beta \epsilon_w = 1.25$ calculated for a series of different values of the fluid-fluid pair potential cut-off range $r_c$. Computer simulations often truncate the particle interactions at $r_c=2.5\sigma$. At this state point, we see here a change in the predicted interfacial phase behaviour as $r_c$ is varied. Even going from $r_c=5\sigma$ to $r_c=10\sigma$ changes the system from {wetting to non-wetting}.}
\label{cont:fig:rangeTrunc}
\end{figure}

In our previous work\cite{hughes15} calculating $g(\Gamma)$ for the LG model, it was found that truncating the particle interactions can have a significant effect on the wetting behaviour. The present continuum DFT results show that this observation is generally true and is not just an artefact of the LG. In Fig.\,\ref{cont:fig:rangeTrunc}, we display the binding potential for $\beta \epsilon=1.1$ and $\beta \epsilon_w = 1.25$ (near to the wetting transition) calculated for a number of different values of the fluid-fluid pair potential cut-off range $r_c$, namely, from $r_c=2.5\sigma$ to $r_c=20\sigma$. In computer simulations it is common to truncate the LG potential at $r_c=2.5\sigma$. However, the results in Fig.\,\ref{cont:fig:rangeTrunc} show that there is even a change in the interfacial phase behaviour from non-wetting to wetting when going from $r_c=5\sigma$ to $r_c=10\sigma$. This shows that the tails of the potentials have a significant effect in determining the wetting behaviour. Fig.\,\ref{cont:fig:rangeTrunc} shows that for this temperature, making the commonly used $r_c=2.5\sigma$ truncation would lead to an incorrect prediction of the wetting behaviour. Note that the binding potentials displayed in Fig.\,\ref{cont:fig:rangeTrunc} are all for the case when the external potential due to the wall has a fixed strength of $\beta \epsilon_w=1.25$ and the range is not truncated. All these binding potentials are for exactly the same point on the bulk fluid phase diagram since we renormalise $\epsilon$ according to the truncation range -- see Eq.\,\eqref{cont:renorm}. This observation is particularly pertinent in light of the good agreement with computer simulation results shown in Table\,\ref{cont:tab:dft-sim}.

\begin{figure}
\includegraphics[width=\columnwidth]{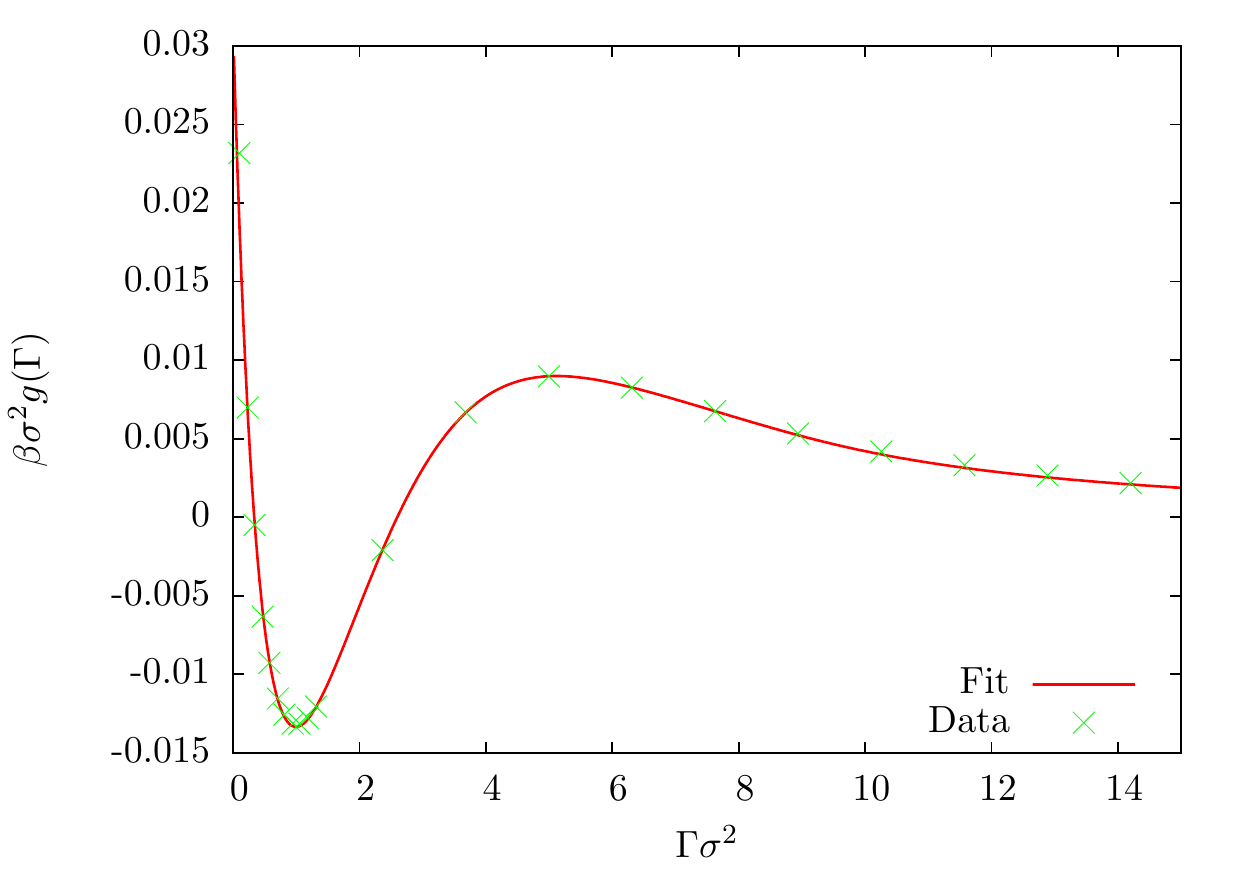}
\caption{A demonstration of the quality of fit achieved with the fit function in Eq.~\eqref{eq:fit}. The binding potential displayed here is for a fluid with $\beta\epsilon = 1.1$, $\beta\epsilon_w = 1.06$ and $r_c = 5${$\sigma$}. The solid line is the fit function and the symbols are the data points from DFT.}
\label{fig:fit}
\end{figure}

\begin{figure}
\includegraphics[width=\columnwidth]{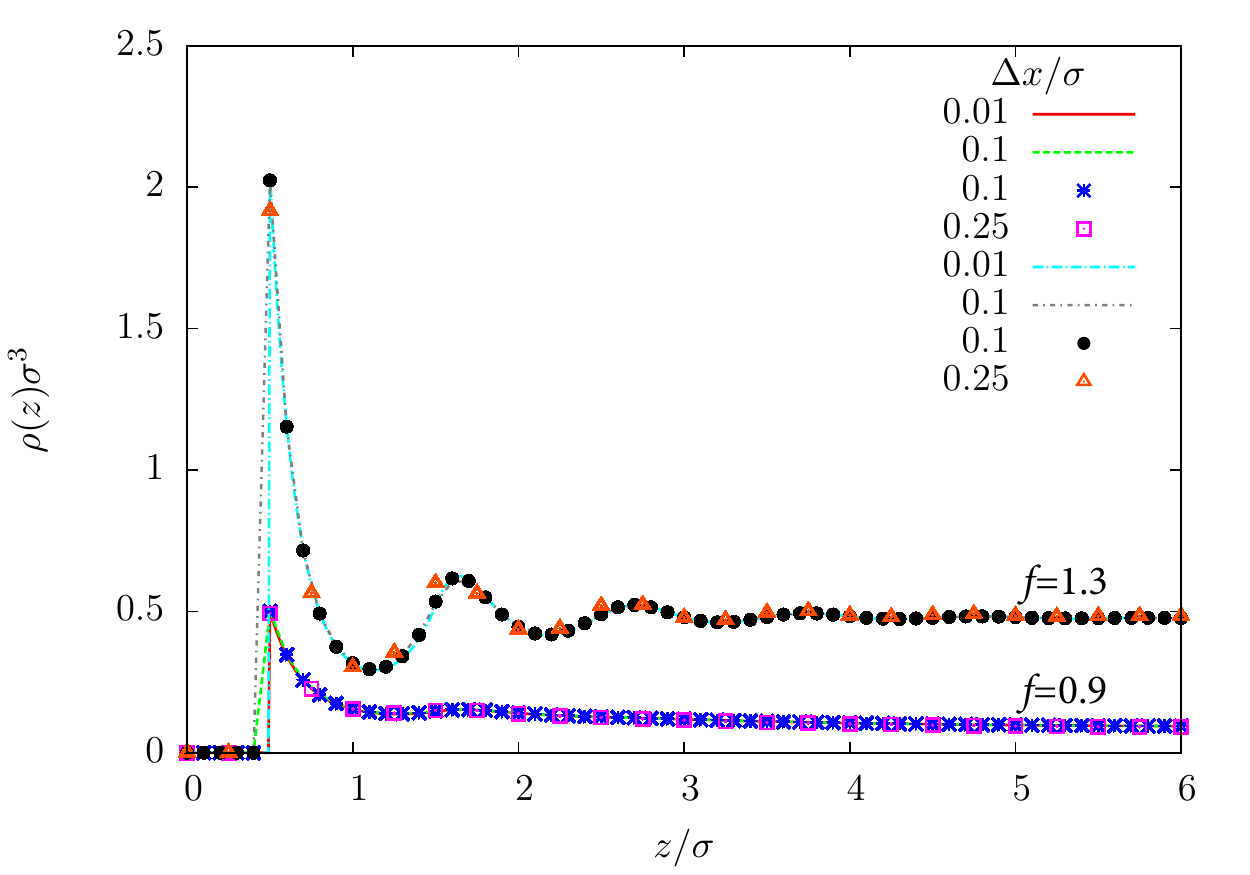}
\caption[Comparison for different dimensionality]{A comparison of equilibrium density profiles obtained using both {the 1D (lines) and 2D (symbols) calculations,} for the fluid with $\beta \epsilon = 1.1$ and $\mu = \mu_{coex}$ against a wall with $f=0.9$ and $f=1.3$. The 1D calculations are performed using a grid spacing of $\Delta x = 0.01 \sigma$ and $\Delta x = 0.1 \sigma$. The 2D profiles are calculated with grid spacings of $\Delta x = 0.1 \sigma$ and $\Delta x = 0.25 \sigma$. The system is non-wetting for $f=0.9$, but a wetting film is observed for $f=1.3$. The profiles calculated using the very coarse grid spacing $\Delta x = 0.25 \sigma$ are in surprisingly good agreement with the finer grid spacings results.}
\label{cont:fig:dimensions}
\end{figure}

Fig.~\ref{fig:fit} illustrates that the form in Eq.\,\eqref{eq:fit} gives an excellent fit to the binding potential for the case when $\beta\epsilon = 1.1$, $\beta\epsilon_w = 1.06$ and $r_c = 5${$\sigma$}. The agreement is just as good for all the other state points discussed so far. Recall that this fit function also gave excellent agreement with the LG results.\cite{hughes15} Some typical values of the fitting parameters are given in the Appendix. 
This fit form for $g(\Gamma)$ can then be used as input to the IH model  [Eq.~\eqref{eq:ih}] to find droplet height profiles and these can also be compared directly with film height profiles extracted from the 2D droplet density profiles calculated using DFT, such as those displayed in Fig.~\ref{cont:fig:2dprof}. Before showing these comparisons, we should remark about an aspect of the numerics used: The binding potential results above are all calculated using a DFT code that assumes that the density profiles only depend on $z$, the perpendicular distance from the wall, and are invariant in the other two directions. In these effective one-dimensional (1D) computations a small grid spacing ($\Delta x = 0.01\sigma$) can be used in the spatial discretisation of the profiles, so that there are no appreciable discretisation errors in any of the quantities calculated. However, when calculating 2D drop profiles of sizes such as those in Fig.~\ref{cont:fig:2dprof} a coarser grid is used (typically $\Delta x = 0.1\sigma$), since otherwise the calculations are just not feasible with the computer resources currently available. The coarser grid can result in small errors, particularly in the integration of the attractive LJ contribution, that can lead to deviations in the comparisons between the contact angles of the two models. These errors are particularly prevalent at state points close to the wetting transition where small shifts in the minimum value of the binding potential, $g(h_0)$, lead to appreciable changes in the calculated contact angle. Therefore, for a consistent comparison, we calculate both the binding potentials and the droplet profiles using the same spatial discretisation and the same 2D code. In Fig.~\ref{cont:fig:dimensions} we display a comparison of density profiles calculated for various values of $\Delta x$. Various grid spacings in both one and two dimensions are displayed and there is good agreement between all of the results. The errors mentioned above occur in quantities obtained via numerical integration involving the density profiles, such as free energies, etc. Given the results in Fig.~\ref{cont:fig:rangeTrunc} showing how sensitive the form of the binding potential is to the range of the pair interactions $r_c$, it is not surprising that there is also a sensitivity to the accuracy of the numerics used in the calculations.

\begin{figure}
\centering
\includegraphics[width=\columnwidth]{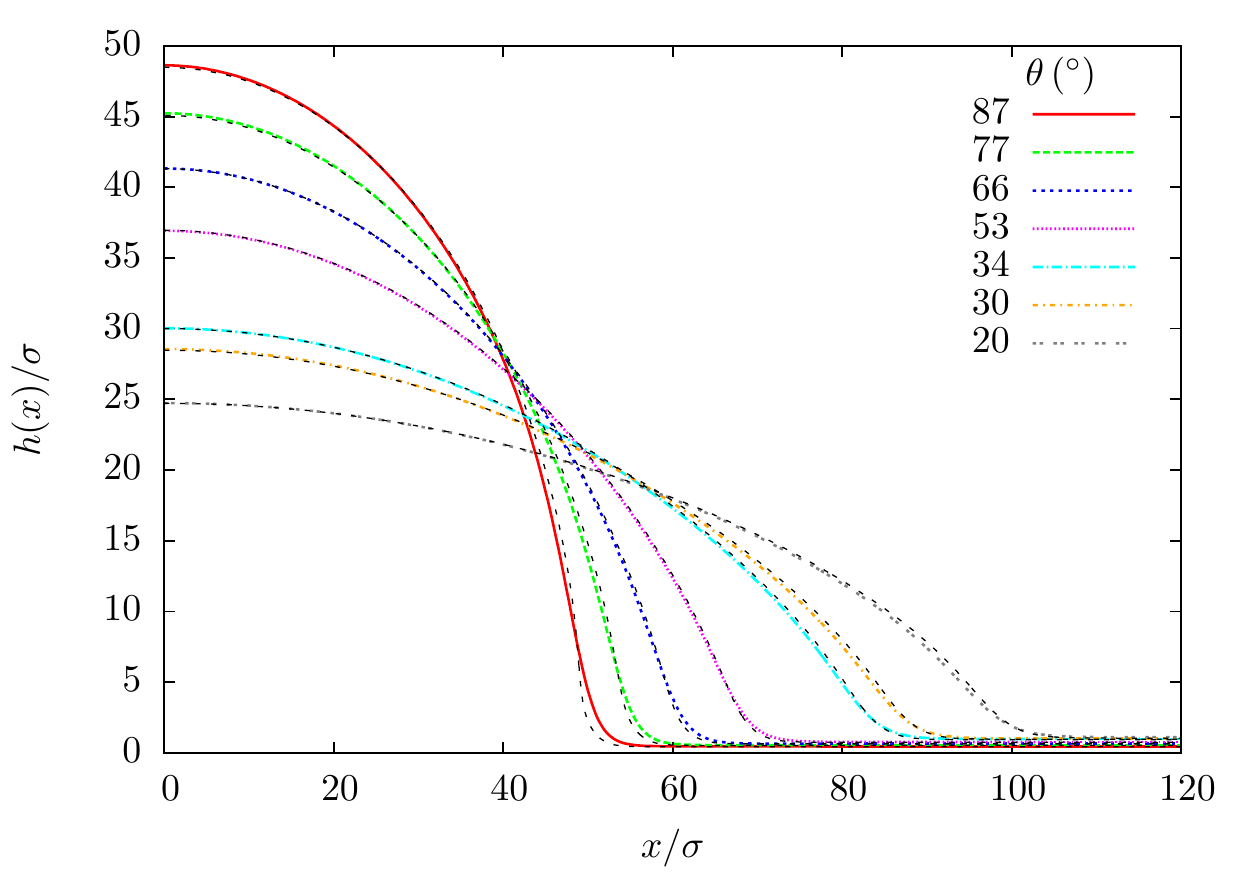}
\caption{A comparison of droplet profiles for the fluid with $\beta \epsilon = 1.1$, $r_c=5${$\sigma$} and $\mu=\mu_{coex}$ with various wall strengths, $f=\ $0.8, 0.85, 0.9, 0.95, 1.01, 1.02, 1.04. The coloured lines show film height profiles calculated from the DFT and the black lines show the corresponding droplet profile found from the IH model  [Eq.~\eqref{eq:ih}] using the binding potential obtained from the DFT as input. Excellent agreement is found for the whole range of contact angles.}
\label{cont:fig:dropComp}
\end{figure}

Fig.\,\ref{cont:fig:dropComp} displays a comparison of the film height profiles obtained from the IH model with the results from DFT for the fluid with $\beta \epsilon = 1.1$ and $r_c=5${$\sigma$} for walls with varying $f$. From the 2D DFT density profiles $\rho(x,z)$, the drop height profile is defined as
\begin{equation}
h(x)=\frac{\Gamma(x)}{(\rho_l-\rho_g)}
\end{equation}
where
\begin{equation}
\Gamma(x)=\int_0^\infty(\rho(x,z)-\rho_g){\rm d}z
\end{equation}
is the local adsorption. The IH model uses as input the binding potential calculated from the DFT. The parameters in the fit function \eqref{eq:fit} are given in the Appendix. Fig.\,\ref{cont:fig:dropComp} shows there is excellent agreement between the two models over a wide range of contact angles. A very small discrepancy exists in the contact line region where the IH model slightly underestimates the film height. The two droplets in each case are constrained to have the same average film height, which is identical to fixing the volume. There is no constraint on the maximum height of the droplets, nevertheless the agreement of this maximum height between the two methods is very good.

\begin{figure}
\includegraphics[width=\columnwidth]{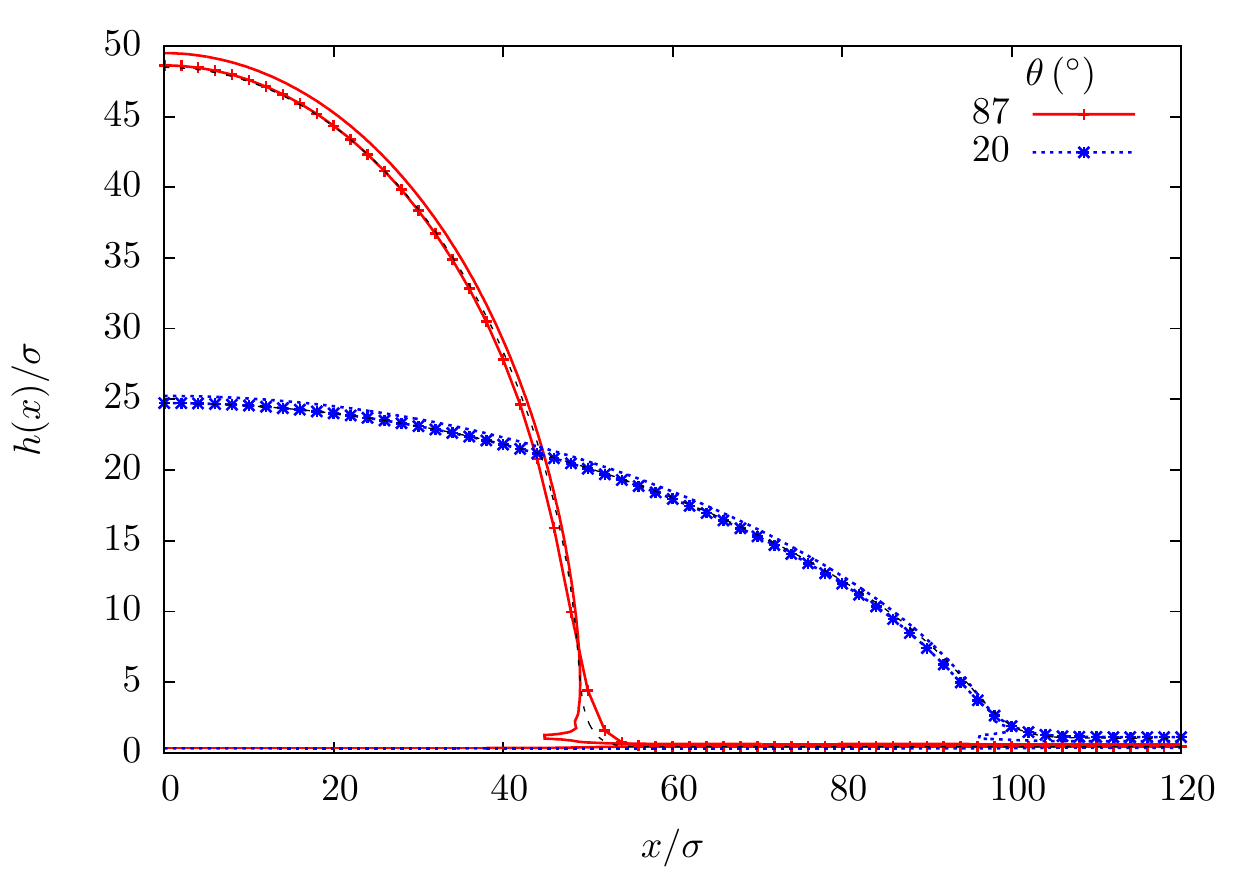}
\caption{A comparison of two methods for extracting a film height profile from a DFT drop density profile, namely, (i) the density contour $\rho(x)=(\rho_l+\rho_g)/2$ (plain lines) and (ii) the height profiles found by calculating the local adsorption (lines with symbols). The black dashed lines show the IH results. These are two cases from Fig.\,\ref{cont:fig:dropComp}, with $f=0.8$ (red solid curve) and $f=1.04$ (blue dotted curve).}
\label{cont:fig:contours}
\end{figure}

An alternative method of finding a film height profile from the density profile of a liquid droplet is to use the contour of a specific density value, such as the density value $(\rho_l+\rho_g)/2$ which is the value half way between the two coexisting densities. Finding a height profile in such a way better shows the structure of the droplets in the contact region, as illustrated in Fig.\,\ref{cont:fig:contours}. However, this does not give as much insight into the nature of the adsorption on the surface away from the droplet. Calculating the local adsorption, illustrated in Fig.\,\ref{cont:fig:dropComp}, clearly allows for better comparison with the IH model and accurately fits the height of the droplet compared to the IH output. It should be noted, however, that for droplets with a contact angle of $\theta>90^\circ$ the local adsorption becomes somewhat misleading and the contour method is a much more meaningful measure of the droplet profile. In such a case only the density contour can accurately describe the droplet, since the adsorption method can not describe the multivalued nature of the droplet profile. 

\begin{figure}
\includegraphics[width=\columnwidth]{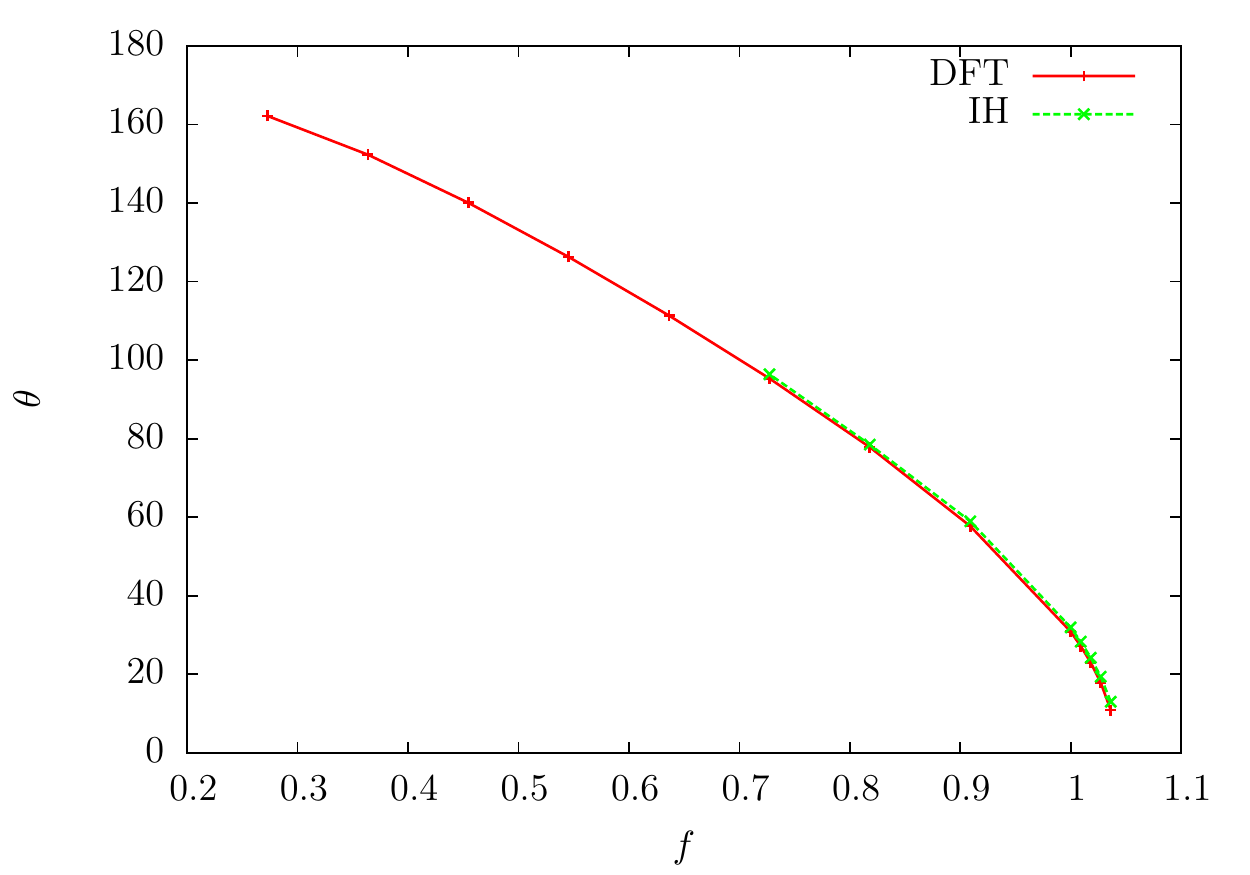}
\caption{A comparison between the contact angle $\theta$ found using DFT and Young's equation \eqref{eq:Young} with the contact angle obtained from drop profiles calculated using the IH model with the binding potential calculated from DFT as input. These results are for varying wall attractive strength $f$ and for fixed $\beta \epsilon = 1.1$ {and $r_c=5\sigma$}.}
\label{cont:fig:contactAngles}
\end{figure}

The contact angle predicted by the two models can be directly compared. Using DFT, the contact angle is obtained  by calculating the three interfacial tensions and then applying Young's equation \eqref{eq:Young}. Using the IH model, the contact angle is extracted from the drop profiles by fitting a circle to the apex of the droplet and finding the contact angle of this circle. In Fig.\,\ref{cont:fig:contactAngles} we see excellent agreement between the two methods. Note that one is only able to extend the DFT results beyond $\theta=90^\circ$. Note too that these results are for the macroscopic contact angle which is only attained by larger droplets. Indeed with DFT one does not have to calculate a droplet profile to calculate $\theta$.

\begin{figure}
\includegraphics[width=\columnwidth]{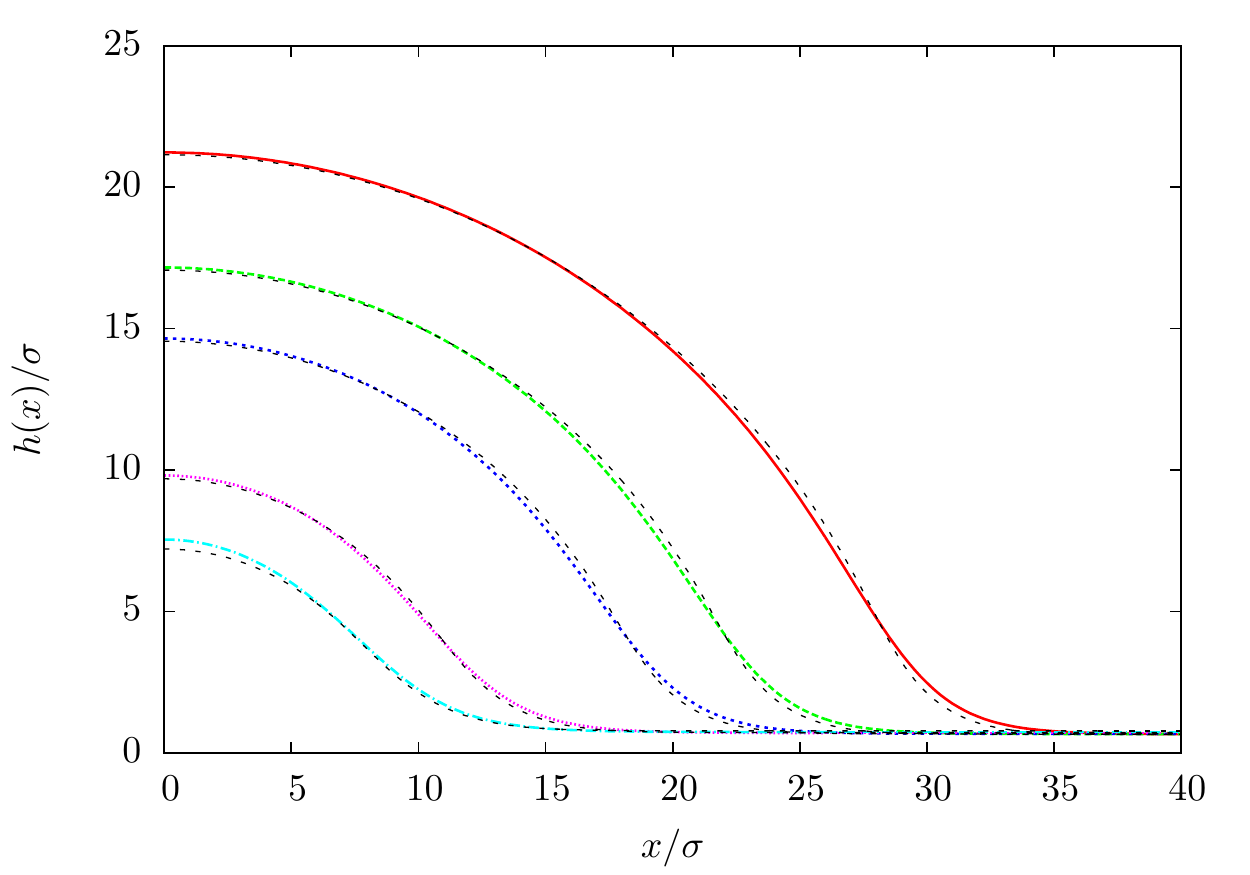}
\caption{Droplet profiles of various volumes when $\beta \epsilon=1.1$ and $f=0.8$. This corresponds to a macroscopic contact angle of $\theta = 87^\circ$. As the volume is varied, good agreement is found between the DFT and IH results, even down to very small droplet sizes.}
\label{cont:fig:dropVol}
\end{figure}

In Fig.\,\ref{cont:fig:dropVol} we display a series of drop profiles as the volume in the drop is varied. We see that the droplet profiles obtained from the two models agree down to very small volumes. As the volume of a droplet increases the contact angle obtained from the profile approaches the macroscopic contact angle. The (coloured) solid lines in Fig.\,\ref{cont:fig:dropComp} show the results from the DFT and the (black) dashed lines are the IH droplet profiles.

We should mention that it is possible to find solutions to the IH model that are droplets with a macroscopic contact angle slightly beyond $\theta=90^\circ$. However, in such cases the agreement with the DFT is not good and fitting the droplet with a circle is a very bad approximation of the film height profile, even though it accurately predicts the macroscopic contact angle. This poor fit stems from the inability of the film height profile to display the multivalued character of the droplet profile.

\section{A model fluid with oscillatory binding potentials}\label{cont:sec:fw}

In the work above for the LJ fluid with pair-potential \eqref{cont:ljts} all of the binding potentials $g(\Gamma)$ decay monotonically with increasing $\Gamma$. One can observe oscillatory binding potentials for this system, however these occur for low temperature states where the liquid phase wetting the wall is expected to be metastable with respect to the crystalline solid phase, so we consider a different model fluid where this is not the case. The occurrence of oscillatory binding potentials is connected to the location of the Fisher-Widom (FW) line,\cite{fisher69} which is the locus in the phase diagram at which the asymptotic decay of the correlations crosses over from monotonic to damped oscillatory. The form of decay can be seen in the bulk fluid radial distribution function ${\cal G}(r)$, given by the Ornstein-Zernike equation,\cite{hansen13} as well as in the decay of the inhomogeneous fluid density profiles. Whether the $r\to\infty$ decay of ${\cal G}(r)$ is monotonic or damped oscillatory is determined by the complex pole with smallest real part that is a solution of $1-\rho_b \hat{c}(k) = 0$, where $\hat{c}(k)$ is the Fourier transform of the bulk fluid pair direct correlation function $c^{(2)}(|\boldr-\boldr'|) = -\beta \delta^2 F_\textrm{ex} / \delta \rho(\boldr)\delta \rho(\boldr')$.\cite{evans93, hansen13}

Whatever the fluid pair potential, there is normally some oscillatory structure at a wall-liquid interface (see e.g., the curve for $f=1.5$ in Fig.\,\ref{cont:fig:1dprof}), even if the ultimate (asymptotic) decay of the density profile is monotonic. However, when the liquid state is sufficiently far from the critical point, the asymptotic decay can instead be oscillatory, depending on which side of the FW that state point lies. The decay from the liquid phase to the gas phase at the liquid-gas interface is almost always monotonic (see Fig.~\ref{cont:fig:constrained1d}(b)).

The fluid we now consider is a colloid-polymer mixture. The effective potential between the colloids in a colloid-polymer mixture has a range that is determined by the radius of gyration of the polymers, $\sigma_p/2$, as well as being determined by the diameter of the colloids, $\sigma$. In these systems, the location of the FW line is closely related to the size ratio $q=\sigma_p /  \sigma$.\cite{brader03} Moreover, in colloid-polymer mixtures one can also find the FW line located well away from freezing which means that in such systems one may observe oscillatory liquid-gas interfacial profiles and multiple layering transitions.\cite{brader2001entropic, brader03} Note that in the colloid-polymer context the ``liquid'' is a colloid rich phase and the ``gas'' is a colloid poor phase. The layering transitions observed in the Asakura-Oosawa (AO) model of colloid-polymer mixtures\cite{brader2001entropic, brader03} are indicative that the binding potential for this system is oscillatory, since each new ``layer'' corresponds to a different minimum in $g(\Gamma)$. Indeed, the presence of layering transitions in any system is indicative of oscillatory binding potentials.

In colloid-polymer mixtures, the effective pair-potential between the colloids can be approximated using the AO potential,\cite{asakura54} which for $\sigma \leq r \leq (1+q)\sigma$,
\begin{equation}\label{cont:aopot}
v_{AO}(r) = -\epsilon \left(1 - \frac{3 r}{2 \sigma(1+q)} + \frac{r^3}{2 \sigma^3(1+q)^3} \right)
\end{equation}
and $v_{AO}(r) = 0$ for $r>{\cal R}$, where ${\cal R}=(1+q) \sigma$. The depth of the attractive well is governed by the parameter
\begin{equation}\label{eq:ao_eps}
\epsilon = \frac{1}{6} \pi \sigma_p^3 z_p \left( \frac{1+q}{q}\right)^3,
\end{equation}
where $z_p$ is the polymer fugacity. Here, we quote simply the value of $\epsilon$; the corresponding value of the fugacity can be obtained via Eq.\ \eqref{eq:ao_eps}. Recall that this attractive potential originates from the entropic gain when colloids are close together, as the excluded volume from the interactions with the polymers overlap.\cite{asakura54} The size ratio $q$ specifies the range ${\cal R}=(1+q) \sigma$ of the potential $v_{AO}(r)$.

Using the functional in Eq.~\eqref{cont:vdw} with $v(r)=v_{AO}(r)$ in Eq.~\eqref{cont:aopot} for $r\geq\sigma$ and $v(r)=v_{AO}(\sigma^+)$ for $r<\sigma$, we calculate the density profile of the colloids in the vicinity of a wall. We assume that the wall is composed of particles in the domain $z<0$ interacting with the fluid particles via the potential in Eq.\,\eqref{cont:aopot}. Thus, as previously, the net interaction of a single fluid particle with the entire wall is found by integrating the pair potential assuming a uniform density distribution of particles in the wall. The parameter governing the net strength of wall potential is $\epsilon_w$. {This approximation for the external potential provides a model in which we can vary the strength of the attraction between the particles and the wall. The effective potential between the colloids and a hard wall in the AO model is given in Ref.\ \onlinecite{brader2001inhomogeneous}.}

\begin{figure}
\includegraphics[width=\columnwidth]{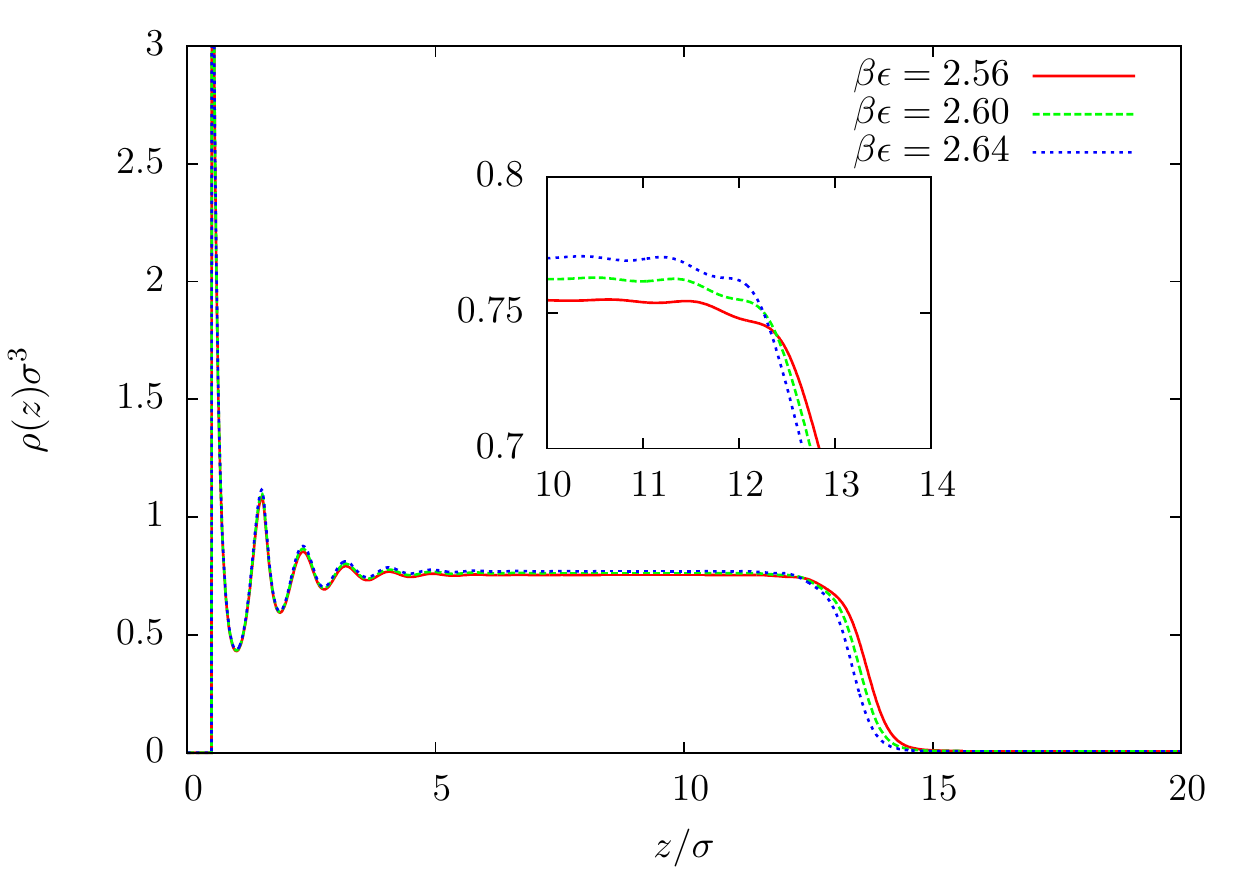}
\caption{Density profiles for the AO fluid with $q=0.5$ and varying $\beta\epsilon$, with the adsorption constrained to be {$\Gamma\sigma^2=10$}. For these values of $\beta\epsilon$ the liquid state on the wall is below the FW line so that the decay of the density profiles into the liquid from both interfaces is oscillatory. Increasing $\beta \epsilon$ leads to the amplitude of these density oscillations increasing and also the amplitude of the oscillations in $g(\Gamma)$. The external potential in all cases has {$\beta \epsilon_w = 3.2$}.}
\label{cont:fig:fwProf}
\end{figure}

In the AO model, when the range of the potential is small (i.e.\ small $q$) and the interaction strength $\epsilon$ is large enough (equivalently, low enough temperature), oscillations can be found on the liquid side of the liquid-gas interface, as can be seen in Fig.\,\ref{cont:fig:fwProf} (see also Refs.~\onlinecite{brader2001entropic, brader03}). The amplitude of these oscillations increases as $\epsilon$ is increased.

\begin{figure}
\includegraphics[width=\columnwidth]{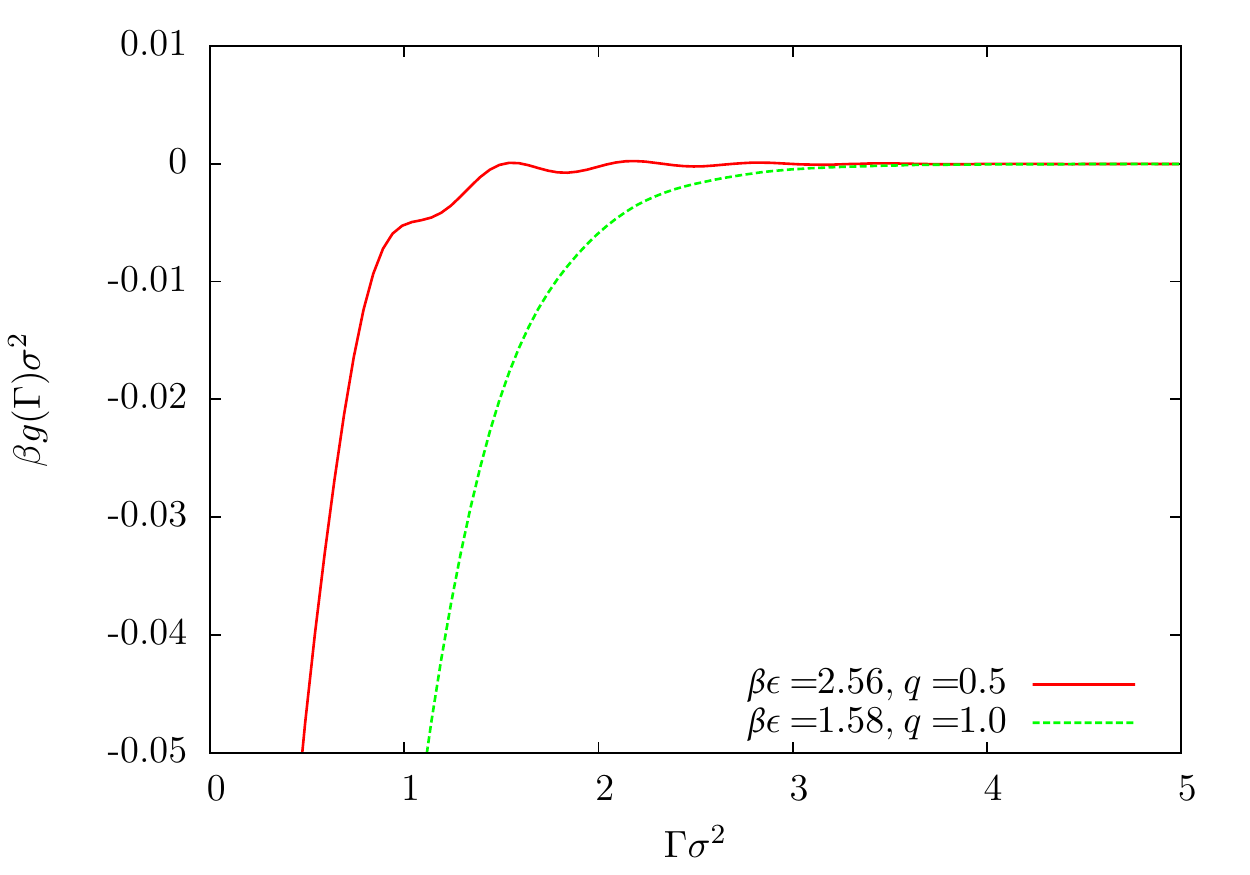}
\caption{The binding potential $g(\Gamma)$ for the AO fluid with two different values of $q$. {The wall attraction strength is $\beta \epsilon_w=2.8$.} The parameters are chosen so that these are at identical state points on the bulk fluid phase diagram. When the range of the pair interactions ${\cal R}=(1+q)\sigma$ is decreased (i.e.\ $q$ is decreased), the range of $g(\Gamma)$ decreases and oscillations appear in the binding potential. {Note that only a portion of the curves is displayed, focusing on the range where oscillations are best seen. See Fig.~\ref{cont:fig:fwBind} for a wider view.}}
\label{cont:fig:fwOnset}
\end{figure}

The binding potentials displayed in Fig.\,\ref{cont:fig:fwOnset} illustrate that the range of the pair potential is a key factor in determining whether or not oscillations occur in the binding potential. The two binding potentials displayed are (within the present mean-field treatment) at the same state point in the bulk fluid phase diagram, since the integrated strength of the pair potential \eqref{cont:aopot} is the same in both cases. However, only the binding potential for the fluid with $q=0.5$ exhibits oscillations, when the pair potential is short ranged, but deeper. In contrast, when $q=1$ the pair-potentials are longer ranged and the binding potential decays monotonically as $\Gamma \to\infty$. This is because the location of the FW line in the bulk fluid phase diagram depends on both the shape and depth of the pair-potential well $\epsilon$, not just on the total integrated interaction strength.

The occurrence of oscillations in $g(\Gamma)$ can also be understood from considering how the wall-liquid and liquid-gas interfaces interact through the liquid. When the interfaces are far apart, the oscillations from each interface decay to the bulk value and so the two interfaces do not interact. As the two interfaces approach each other the envelopes of the oscillations from each interface overlap. In effect, the liquid gas interface constrains the shape of the density oscillations from the wall-liquid interface and similarly, the wall-liquid interfaces constrains the liquid-gas interface. Any constraint must raise the free energy of the system. Some film thicknesses (i.e.~$\Gamma$ values) raise the free energy more than others because the oscillations from one interface can not smoothly transition into the oscillations of the other. This leads to oscillations in the free energy $g(\Gamma)$.

\begin{figure}
\includegraphics[width=\columnwidth]{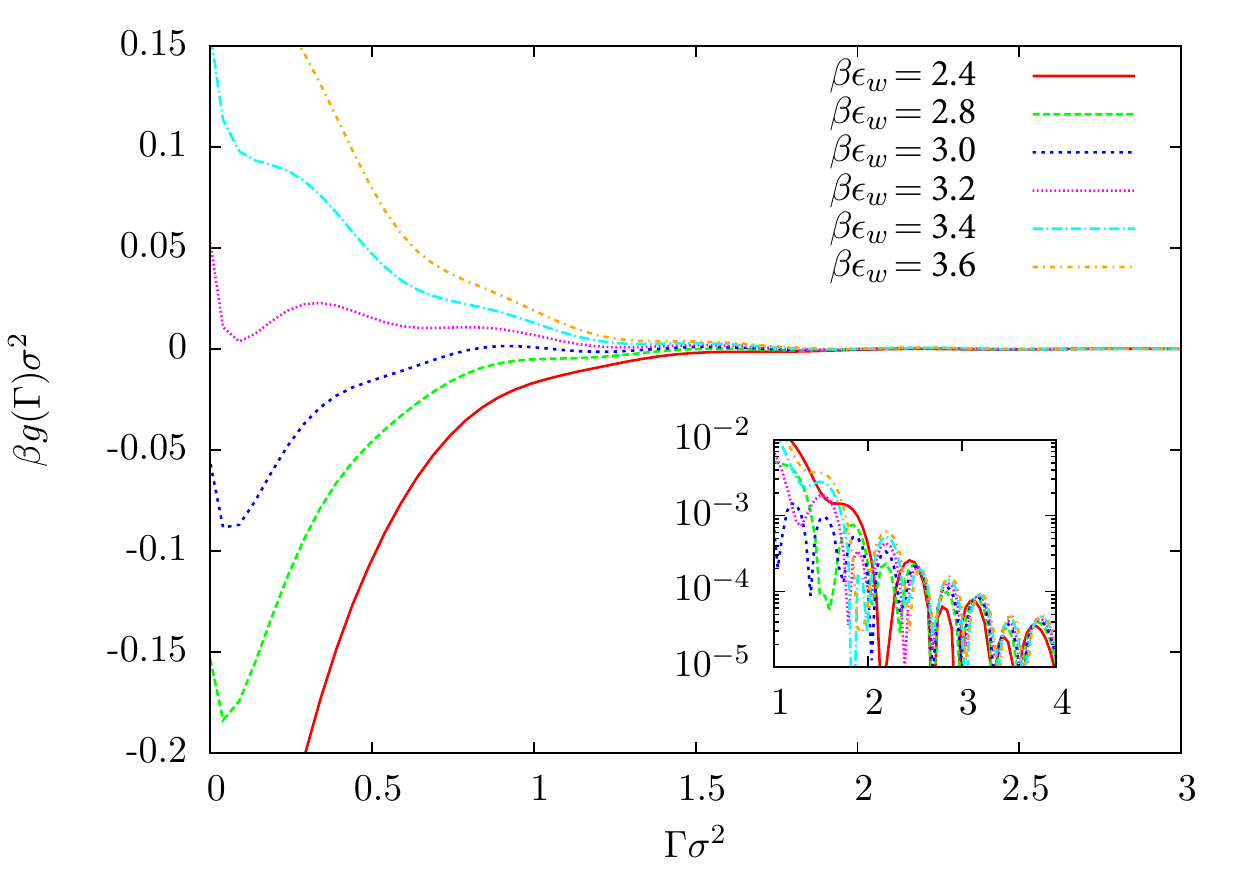}
\caption{The binding potential $g(\Gamma)$ for the AO fluid with $\beta \epsilon = 2.56$ and $q = 0.5$ as the wall attraction strength $\epsilon_w$ is varied. Oscillations are present in $g(\Gamma)$ due to the oscillations in the density profiles. The oscillatory structure from the wall-liquid interface can not smoothly transition into the oscillations of the liquid-gas interface at all film heights. The inset shows $|\sigma^2\beta g(\Gamma)|$ and the logarithmic scale makes the oscillatory decay clearly visible.}
\label{cont:fig:fwBind}
\end{figure}

Fig.\,\ref{cont:fig:fwBind} displays the binding potential for the AO fluid with $\beta \epsilon = 2.56$ and $q = 0.5$ as the wall attraction strength $\epsilon_w$ is varied. As before, when $\epsilon_w$ is increased, the system transitions from non-wetting to wetting. However, now we see the presence of multiple minima in $g(\Gamma)$ for values of $\epsilon_w$ near to the wetting transition. These minima are due to the oscillations in $g(\Gamma)$ and so give rise to the multiple layering transitions studied in detail by Brader and co-workers.\cite{brader2001entropic, brader03}

\begin{figure}
\includegraphics[width=0.99\columnwidth]{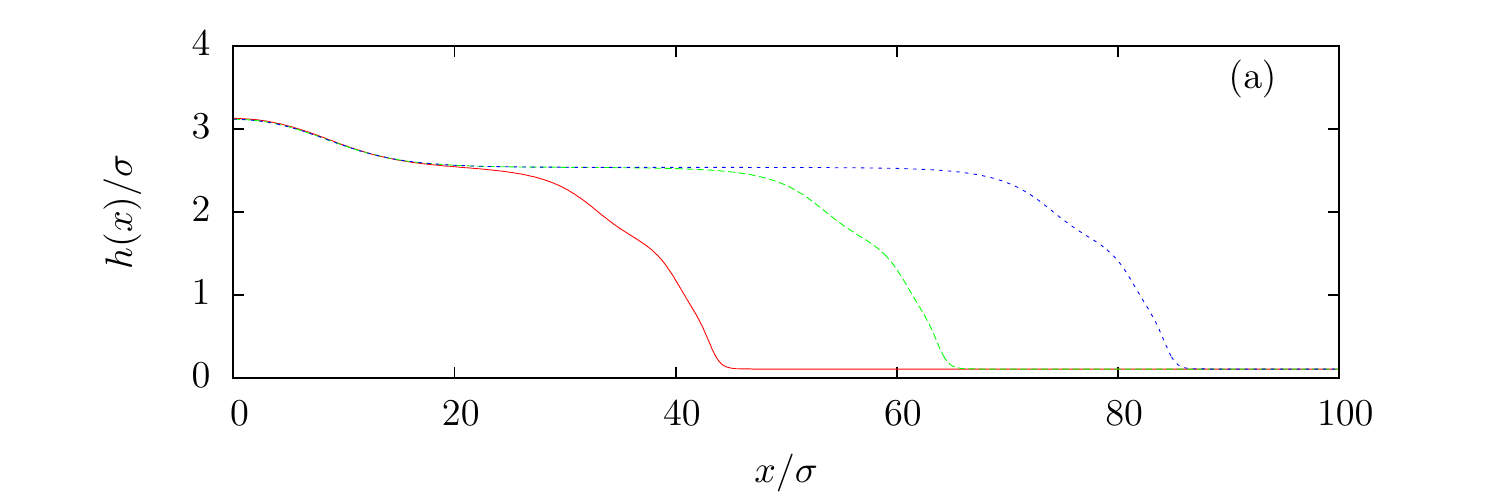}

\includegraphics[width=0.95\columnwidth]{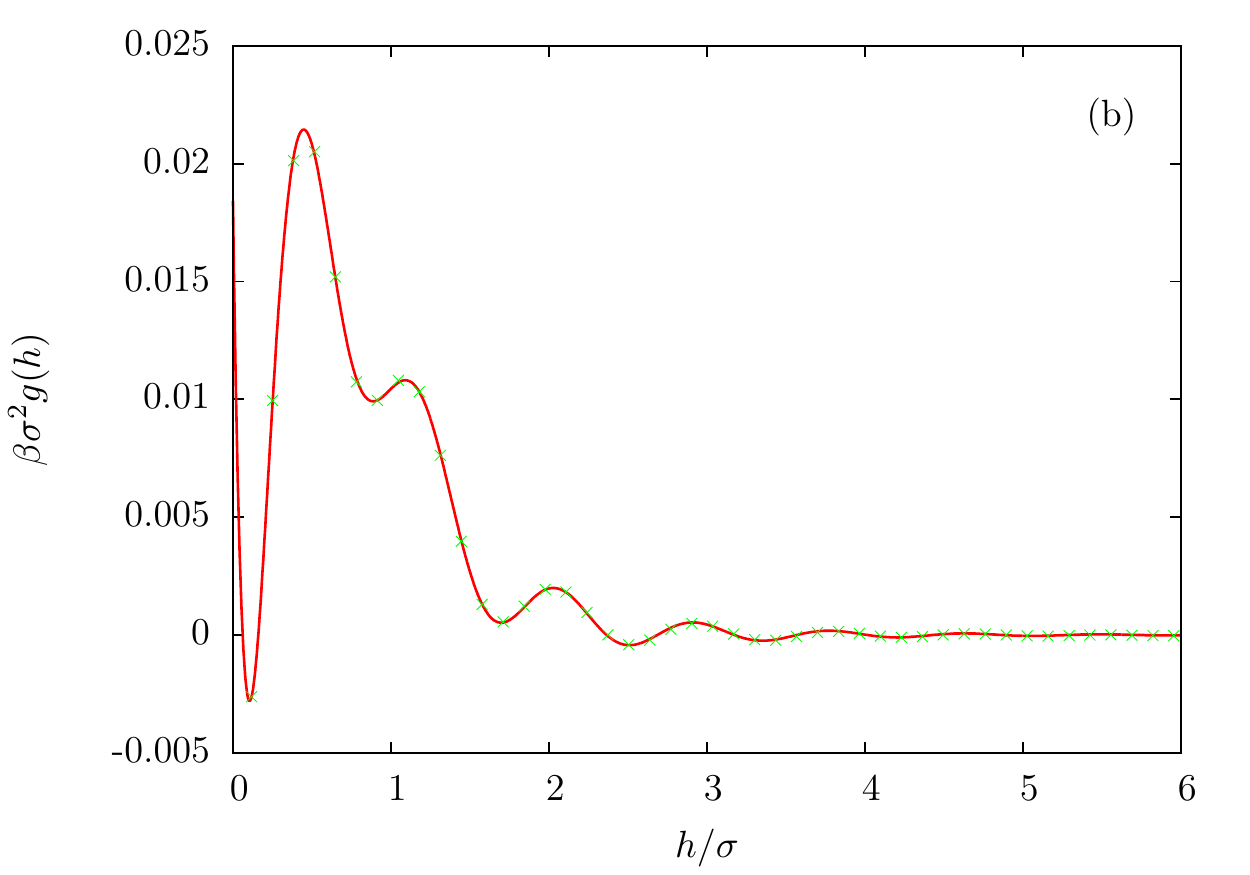}

\caption{In (a) we display film height profiles for different drop volumes calculated from the IH model together with the oscillatory binding potential in Eq.~\eqref{cont:oscilFit}, displayed in (b). The droplet forms distinct layers and spreads out as the volume is increased with no increase in the maximum height. These results are for the AO fluid with $\beta \epsilon=2.6$ and $\beta \epsilon_w=3.24$. The parameters in the binding potential fit Eq.~\eqref{cont:oscilFit} are: $a_0=0.908$, $a_1=-7.352$, $a_2=5.901$, $a_3=-0.011$, $a_4=-0.00015$, $a_5=0.045$, $a_6=0.423$, $a_7=-0.77$, $a_8=-0.231$, $a_9=0.559$.}
\label{cont:fig:oscilDrop}
\end{figure}

The presence of the oscillations in $g(\Gamma)$ means that the fit function in Eq.\,\eqref{eq:fit} is no longer appropriate. For the present system with only short-range interactions (i.e.\ no London dispersion interactions) we find that the following form gives a good fit to the DFT data:
\begin{eqnarray}\label{cont:oscilFit}
\hspace{-6mm}g_f(h) =  a_3 \cos(a_1 h + a_2) e^{-h/a_0}  +  \sum_{n=1}^{m-3} a_{n+3} e^{-nh/a_0},
\end{eqnarray}
where $a_0$, $a_1, \cdots, a_m$ are parameters to be fitted. Typically, we go to $m=9$, but less is often acceptable. The first term accounts for the oscillatory behaviour. The remaining terms are what one finds when only short range forces are present and the decay of the density profiles is monotonic.\cite{dietrich88, schick90} Using this fit function as an input to the IH model, droplet profiles such as those shown in Fig.\,\ref{cont:fig:oscilDrop} are obtained. Panel (a) shows three droplets of different volumes, all obtained with the binding potential displayed in panel~(b). The droplets are very thin and clearly show a layering structure in the contact region. The binding potential in (b) is given as a function of film height using Eq.~\eqref{eq:adsheight}. Note that the minima  of $g(h)$ correspond to the observed steps in the droplet profiles.

\begin{figure}

\includegraphics[width=0.9\columnwidth]{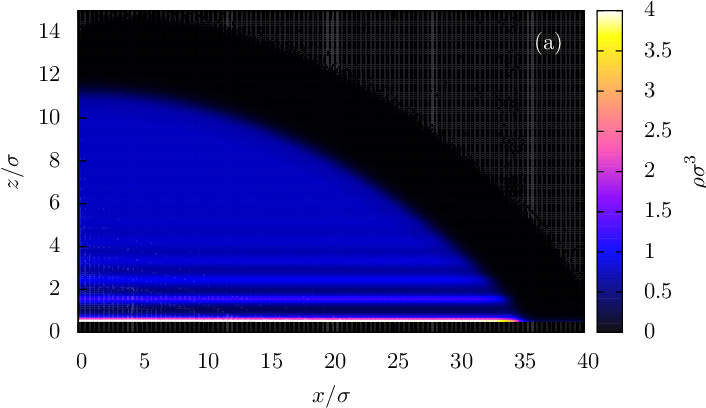}

\includegraphics[width=0.8\columnwidth]{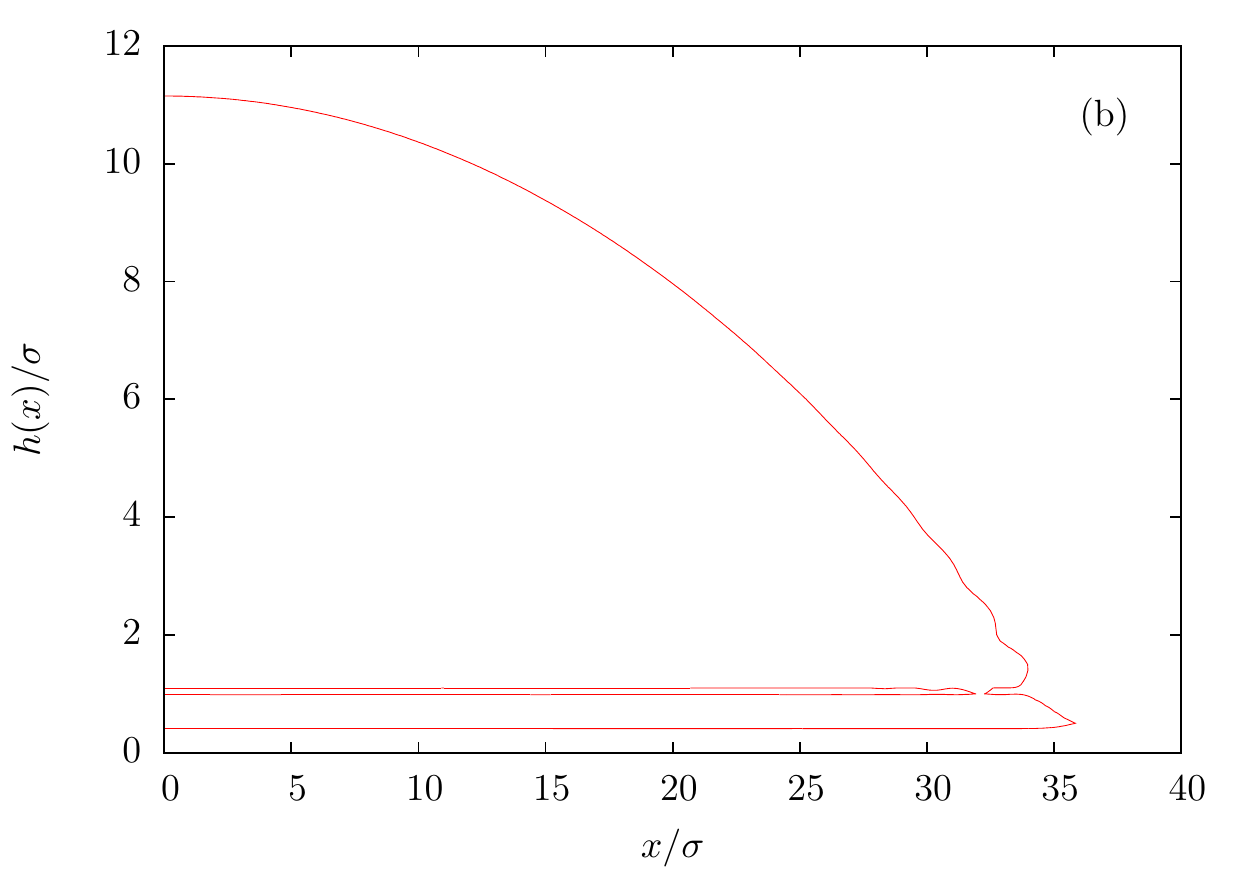}

\includegraphics[width=0.9\columnwidth]{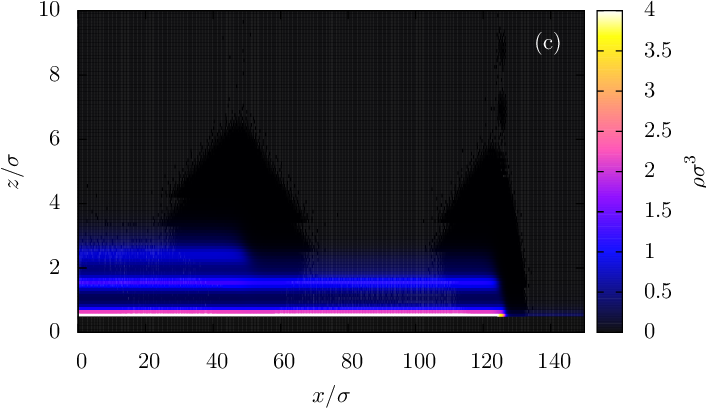}

\includegraphics[width=0.9\columnwidth]{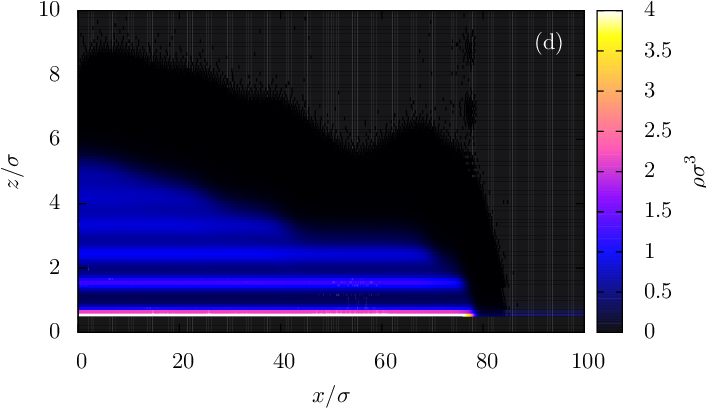}

\caption{(a) and (c) are equilibrium droplet density profiles for $\beta \epsilon=2.6$ and $\beta \epsilon_w=3.24$. The significant structure in the drops, particularly in the contact region, can also be clearly seen from the $\rho \sigma^3=0.375$ contour in (b), which corresponds to the profile in (a). The non-equilibrium intermediate state in (d) points to the possibility of novel spreading dynamics.}
\label{cont:fig:dftOscilDrop}
\end{figure}

The droplet profiles can also be found directly using DFT as shown in Fig.\,\ref{cont:fig:dftOscilDrop}, although, of course, the DFT gives a much more detailed description of the structure on the surface. The drop profiles in Fig.\,\ref{cont:fig:dftOscilDrop} exhibit much more structure than the droplets displayed in Fig.~\ref{cont:fig:2dprof} which are for state points on the monotonic-decay side of the FW line. From the binding potential in Fig.~\ref{cont:fig:oscilDrop} we see that the drop in Fig.\,\ref{cont:fig:dftOscilDrop}(c) is for a state point near to the wetting transition. In the contact line region of these drops a terrace-like structure is found. This is most clearly observed in the $\rho \sigma^3=0.375$ contour displayed in Fig.\,\ref{cont:fig:dftOscilDrop}(b). Above this stepped region the droplet takes on the form of a spherical cap. On approaching the wetting transition the droplets are more spread out, becoming very wide and flat, as shown in Fig.\,\ref{cont:fig:dftOscilDrop}(c). Here there is no spherical cap component to the droplet and instead it is very flat, dominated by the local minima of the oscillations of the binding potential. Each minimum corresponds to having an additional layer of particles at the wall. Displayed in Fig.\,\ref{cont:fig:oscilDrop}(b), the global minimum in $g(h)$ is at $h\approx 0.1\sigma$, which corresponds to a small, sub-monolayer, number of particles adsorbed on the wall. This can also be seen from the density profiles outside the droplets in Fig.\,\ref{cont:fig:dftOscilDrop}. Increasing the film thickness, the next local minimum in $g(h)$ in Fig.\,\ref{cont:fig:oscilDrop} is at $h \approx \sigma$, corresponding to an almost complete monolayer adsorbed on the wall. However, the free energy for this configuration is much higher than the ones for the two minima either side, corresponding to a near empty surface ($h \approx 0.1\sigma$) and two adsorbed layers ($h \approx 1.7\sigma$), respectively. The minimum in $g(h)$ at $h \approx 2.5\sigma$, corresponding to three layers of particles, is an even lower free energy state. The fact that two or three almost complete layers are  lower energy states than a single layer is reflected in the density profiles in Fig.\,\ref{cont:fig:dftOscilDrop}, where a single layer of particles adsorbed at the wall can not be observed.

Finally, Fig.\,\ref{cont:fig:dftOscilDrop}(d) shows an intermediary {\em non-equilibrium} density profile found during the minimisation to the droplet shown in (c). The initial condition was half a circular disc of the liquid density surrounded by the gas. As the Picard iterations proceed, the drop spreads by first advancing through a `precursor' film composed of two layers of colloids. It must be stressed that this iterative minimisation procedure is not necessarily representative of the true spreading dynamics of the system, {but may still be considered as representing a ``pseudo-dynamics''. Then, intermediary forms such as that in (d) suggest that the evolution of these droplets towards equilibrium may correspond to very interesting spreading dynamics. We mention that in some other systems \cite{archer15} the pseudo-dynamics generated by the Picard iteration is actually very similar to the true dynamics from dynamical density functional theory (DDFT).\cite{marconi1999dynamic, marconi2000dynamic, AE04, AR04} Here, the spreading dynamics is not discussed further; this is pursued in Ref.~\onlinecite{yin2016films}.}

\section{Summary and conclusions}\label{sec:conc}

We have presented results from a DFT based method for calculating the binding potential, which is based on calculating density profiles with an adsorption constraint. Comparing drop height (adsorption) profiles from solving the DFT in 2D with those from using the obtained binding potential in the IH model in Eq.~\eqref{eq:ih}, shows good agreement, validating this coarse-graining procedure. The liquids considered consist of particles interacting via a hard-sphere potential and an additional attraction that is treated in a mean field fashion.\cite{evans79,evans92} The reference hard-sphere system is described using the accurate White Bear version of FMT.\cite{hansen13, roth02, roth10} The first case considered has attraction given by a truncated and shifted Lennard-Jones potential. Comparing with molecular dynamics simulation results indicates the DFT describes the system fairly well -- see Table\,\ref{cont:tab:dft-sim}. 

Our results demonstrate that truncating fluid-fluid pair interactions can have profound effects on the predicted wetting behaviour. Specifically, it was shown that at certain state points (see Fig.~\ref{cont:fig:rangeTrunc}), changing the truncation range even from $r_c=5\sigma$ to $r_c=10\sigma$, leads to a change in the predicted interfacial phase behaviour, from wetting to non-wetting. This shift in phase behaviour occurs far beyond the usual truncation range of $r_c=2.5\sigma$ that is used in most computer simulations.

To input the calculated binding potentials into the IH model to calculate droplet profiles, they are fitted to an algebraic form. These droplet profiles can then be compared to droplet profiles calculated directly using DFT by integrating over the 2D density profile to obtain the film height profile. Excellent agreement is found between the two methods, including even for small droplet volumes. Comparison of macroscopic contact angles from each model also showed excellent agreement is found for all contact angles $0^\circ < \theta < 90^\circ$. However, only the DFT is able to find droplet profiles for $\theta > 90^\circ$.

The AO model pair potential has also been considered, being a typical model system with a short-range attractive pair interaction potential, compared to the diameter of the particles. For this system, at lower effective temperatures (higher $\epsilon$), density profiles with oscillatory decay into the liquid are observed. These occur when the bulk liquid state is to the right of the FW line in the phase diagram. At such state points, there can also be oscillations on the liquid side of the density profile at the liquid-gas interface.\cite{brader2001entropic, brader03} Of course, there are also oscillations in the liquid density profile near the wall. These oscillatory density profiles give rise to oscillations in the tails of the binding potentials. By fitting the algebraic form in Eq.\,\eqref{cont:oscilFit}, suitable for short ranged interactions, to these binding potentials, droplet height profiles can be found from the IH model. The droplet profiles for state points close to the wetting transition have a very pronounced terraced structure, are very thin and can lose their spherical cap shape. Such droplet profiles can also be found directly using the DFT and exhibit the same stepped structure. A spherical cap is still found for larger droplets but there remains a high degree of structuring in the contact region of the drops. 

The droplet profiles observed in Figs.\,\ref{cont:fig:oscilDrop} and \ref{cont:fig:dftOscilDrop} are remarkably similar to droplet profiles observed in the experiments reported in Ref.\,\onlinecite{heslot89}. See also the discussions in Refs.\,\onlinecite{de2013capillarity, popescu12, KrDe1995cpl, DeCh1974jcis, WaNi2003n, CNWT2004jcis, matar2007dynamic, hu2012influences}. The experiments in Ref.\,\onlinecite{heslot89} were for PDMS on a silicon wafer surface and terraced droplet profiles were observed, pointing to an underlying oscillatory binding potential. Layered and terraced droplets have also been observed in atomistic MD simulations of small droplets on surfaces.\cite{yang1992terraced, bekink96, tretyakov13, isele2016requirements} The striking non-equilibrium droplet profile, with a `precursor-film' advancing in front of the droplet displayed in Fig.~\ref{cont:fig:dftOscilDrop}(d), {found during the pseudo-dynamics formed by the} iterative minimisation procedure used to calculate the density profiles, indicates that this system is capable of exhibiting unusual spreading dynamics like that observed experimentally in Ref.\ \onlinecite{heslot89}. Such spreading dynamics is the subject of a separate study,\cite{yin2016films} based on Eq.\ \eqref{eq:tfe}.

{Finally, we should mention that although the local IH \eqref{eq:ih} with binding potential $g$ obtained from DFT as input is sufficient for the cases studied here, Parry and co-workers have showed that the true effective IH is in fact non-local.\cite{parry06, parry07, bernardino09, fernandez2013intrinsic} At wetting transitions non-locality and fluctuation effects are particularly important. This is of particular relevance to the AO model fluid discussed in Sec.\ \ref{cont:sec:fw}, due to the short-ranged interactions. Going beyond the present mean field treatment, we expect capillary-wave-like fluctuations to somewhat smear out the terraces in the profile $h(\boldx)$, particularly away from the surface.}

\section{Appendix}

Several binding potentials are used in the results throughout this paper. The parameters used to fit this data to an analytic fitting function are given in Table \ref{apptab:parameters}. The fit function is also repeated here for convenience:
\begin{equation}\label{appeq:fitFunc}
g(\Gamma) = {{\cal A}} \frac{\exp[- P(\Gamma)] -1}{\Gamma^2},
\end{equation}
with
\begin{equation}
P(\Gamma) = a_0 \Gamma^2 e^{-a_1 \Gamma} +a_2 \Gamma^2 + a_3 \Gamma^3 + a_4\Gamma^4 + a_5\Gamma^5.	
\end{equation}

\begin{table}[h]
\begin{tabular}{>{$}c<{$}>{$}c<{$}>{$}c<{$}>{$}c<{$}>{$}c<{$}>{$}c<{$}>{$}c<{$}>{$}c<{$}>{$}c<{$}>{$}c<{$}}
\hline
\beta \epsilon & f & r_c & A & a_0 & a_1 & a_2 & a_3 & a_4 & a_5 \\[1em] \hline
1.1 & 0.8 & 5 & -0.310 & 0.147 & 3.621 & -0.378 & 0.174 & -0.028 & 0.002 \\
1.1 & 0.85 & 5 & -0.336 & 0.151 & 3.237 & -0.302 & 0.139 & -0.021 & 0.001 \\
1.1 & 0.9 & 5 & -0.362 & 0.155 & 2.962 & -0.233 & 0.109 & -0.016 & 0.0008 \\
1.1 & 0.95 & 5 & -0.388 & 0.158 & 2.774 & -0.168 & 0.081 & -0.012 & 0.0006 \\
1.1 & 1.01 & 5 & -0.420 & 0.160 & 2.663 & -0.093 & 0.051 & -0.007 & 0.0004 \\
1.1 & 1.02 & 5 & -0.426 & 0.160 & 2.658 & -0.081 & 0.043 & -0.006 & 0.0003 \\
1.1 & 1.04 & 5 & -0.437 & 0.160 & 2.659 & -0.056 & 0.036 & -0.005 & 0.0002 \\
\hline
\end{tabular}
\caption{Parameter values from fitting the binding potentials calculated using DFT to the fit function given in Eq.\,\eqref{appeq:fitFunc}.}
\label{apptab:parameters}
\end{table}


%

\end{document}